\documentclass[11pt,onecolumn]{IEEEtran}

\usepackage{graphicx}
\usepackage{color}
\usepackage{amsfonts}
\usepackage{cite}
\usepackage[margin=1in]{geometry}
\usepackage{fancyhdr}
\usepackage[bvect]{commands10}
\usepackage{verbatim}

\newtheorem{theorem}{Theorem}
\newtheorem{lemma}{Lemma}
\newtheorem{corollary}{Corollary}
\newtheorem{proposition}{Proposition}
\newtheorem{conjecture}{Conjecture}

\def\myfigwidth{3.25in}

\renewcommand{\expb}[1]{ \exp \left( #1 \right) } 
\newcommand{\snr}{{ \mathsf{snr} }} 
\newcommand{\mmse}{ {\mathsf{mmse}} } 
\newcommand{\logpr}[1]{ \log \left( {#1} \right) } 

\usepackage{amssymb}
\usepackage{prettyref}
\newrefformat{s}{Section~\ref{#1}}
\newrefformat{f}{Fig.~\ref{#1}}
\newrefformat{lmm}{Lemma~\ref{#1}}
\newrefformat{pr}{Proposition~\ref{#1}}
\newrefformat{a}{Appendix~\ref{#1}}
\newrefformat{cr}{Corollary~\ref{#1}}
\newrefformat{cj}{Conjecture~\ref{#1}}

\newcommand{\Bigo}[1]{{{\rm O}\left(#1\right)}}

\newcommand{\fracp}[2]{\frac{\partial #1}{\partial #2}}
\newcommand{\fracpk}[3]{\frac{\partial^{#3} #1}{\partial #2^{#3}}}
\newcommand{\fracd}[2]{\frac{\diff #1}{\diff #2}}
\newcommand{\fracdk}[3]{\frac{\diff^{#3} #1}{\diff #2^{#3}}}
\newcommand{\Rerm}[1]{{\rm Re}(#1)}

\newcommand{\pth}[1]{\left( #1 \right)}

\begin{document}

\author{Dongning Guo, 
Yihong Wu,
  Shlomo Shamai (Shitz), 
  and Sergio Verd\'u 
  \def\thefootnote{}
  \thanks{%
    This work has been partially supported by NSF under grants
    CCF-0644344 and CCF-0635154, DARPA under grant W911NF-07-1-0028,
    and the Binational US-Israel Scientific Foundation.
  }
  \thanks{D.~Guo is with the Department of Electrical Engineering and
    Computer Science, Northwestern University, Evanston, IL 60208,
    USA.  
  } \thanks{Y.~Wu and S.~Verd\'u are with the Department of Electrical
    Engineering, Princeton University, Princeton, NJ 08544, USA.
  }
  \thanks{S.~Shamai (Shitz) is with the Department of Electrical
    Engineering, Technion-Israel Institute of Technology, 32000 Haifa,
    Israel.  
  } 
}

\IEEEoverridecommandlockouts 

\title{Estimation in Gaussian Noise: Properties of the Minimum
  Mean-Square Error}

\maketitle

\thispagestyle{fancy}




\begin{abstract}
  Consider the minimum mean-square error (MMSE) of estimating an
  arbitrary random variable from its observation contaminated by
  Gaussian noise.  The MMSE can be regarded as a function of the
  signal-to-noise ratio (SNR) as well as a functional of the input
  distribution (of the random variable to be estimated).
  It is shown that the MMSE is concave in the input distribution at
  any given SNR.  For a given input distribution, the MMSE is found to
  be infinitely differentiable at all positive SNR, and in fact a real
  analytic function in SNR
  under mild conditions.  The key to
  these regularity results is that the posterior distribution
  conditioned on the observation through Gaussian channels always
  decays at least as quickly as some Gaussian density.  Furthermore,
  simple expressions for the first three derivatives of the MMSE with
  respect to the SNR are obtained.
  It is also shown that, as functions of the SNR, the curves for the
  MMSE of a Gaussian input and that of a non-Gaussian input cross at
  most once over all SNRs.  These properties lead to simple proofs of
  the facts that Gaussian inputs achieve both the secrecy capacity of
  scalar Gaussian wiretap channels and the capacity of scalar Gaussian
  broadcast channels, as well as a simple proof of the entropy power
  inequality in the special case where one of the variables is
  Gaussian.
\end{abstract}

\noindent{\bf Index Terms:} 
Entropy, estimation, Gaussian noise, Gaussian broadcast channel,
Gaussian wiretap channel, minimum mean-square error (MMSE), mutual
information.

\section{Introduction}
\label{s:int}

The concept of mean-square error has assumed a central role in the
theory and practice of estimation since the time of Gauss and
Legendre.  In particular, minimization of mean-square error underlies
numerous methods in statistical sciences.
The focus of this paper is the minimum mean-square error (MMSE) of
estimating an arbitrary random variable contaminated by additive
Gaussian noise.
  
Let $(X,Y)$ be random variables with arbitrary joint distribution.
Throughout the paper, $\expect{\cdot}$ denotes the expectation with
respect to the joint distribution of all random variables in the
braces, and $\expcnd{X}{Y}$ denotes
the conditional mean estimate of $X$ given $Y$.
The corresponding conditional variance is a
function of $Y$ which is denote by
\begin{equation}  \label{eq:cv}
  \var{X|Y} = \expcnd{ (X-\expcnd{X}{Y})^2 }{ Y }.
\end{equation}
It is well known that the conditional mean estimate is optimal in the
mean-square sense.  In fact, the MMSE of estimating $X$ given $Y$ is
nothing but the average conditional variance:
\begin{equation}  \label{eq:mmsevar}
  \mmse(X|Y) = \expect{ \var{X|Y} }.
\end{equation}

In this paper, we are mainly interested in random variables related
through models of the following form:
\begin{equation} \label{eq:yxn}
  Y = \sqrt{\snr} \, X + N
\end{equation}
where $N\sim \mathcal{N} (0,1)$ is standard Gaussian throughout this
paper unless otherwise stated.
The MMSE of estimating the {\em input} $X$ of the model given the
noisy {\em output} $Y$ is alternatively denoted by:
\begin{align} \label{eq:mmse}
  \mmse(X,\snr) 
  &= \mmse\left( X | \sqrt{\snr}\,X+N \right) \\
  &= \expect{ \left( X\! -\! \expect{X|\sqrt{\snr}\,X+N} \right)^2 }.
\end{align}

The MMSE~\eqref{eq:mmse} can be regarded as a function of the
signal-to-noise ratio (SNR) for every given distribution $P_X$, and as
a functional of the input distribution $P_X$ for every given SNR.
In particular, for a Gaussian input with mean $m$ and variance
$\sigma_X^2$, denoted by $X\sim\mathcal{N} \left(m,
  \sigma_X^2\right)$,
\begin{equation}
  \mmse(X,\snr) = \frac{\sigma_X^2}{1+\sigma_X^2\snr}\ .
\end{equation}
If $X$ is equally likely to take $\pm1$, then
\begin{equation} \label{eq:mmseb}
  \mmse(X,\snr) = 1 - \int^\infty_{-\infty}
  \frac{ e^{-\frac{y^2}{2}} }{ \sqrt{2\pi} }
  \tanh( \snr - \sqrt{\snr}\, y ) \,\diff y \ .
\end{equation}
\begin{figure} 
  \small{
  \centering
  \includegraphics[width=.55\columnwidth]{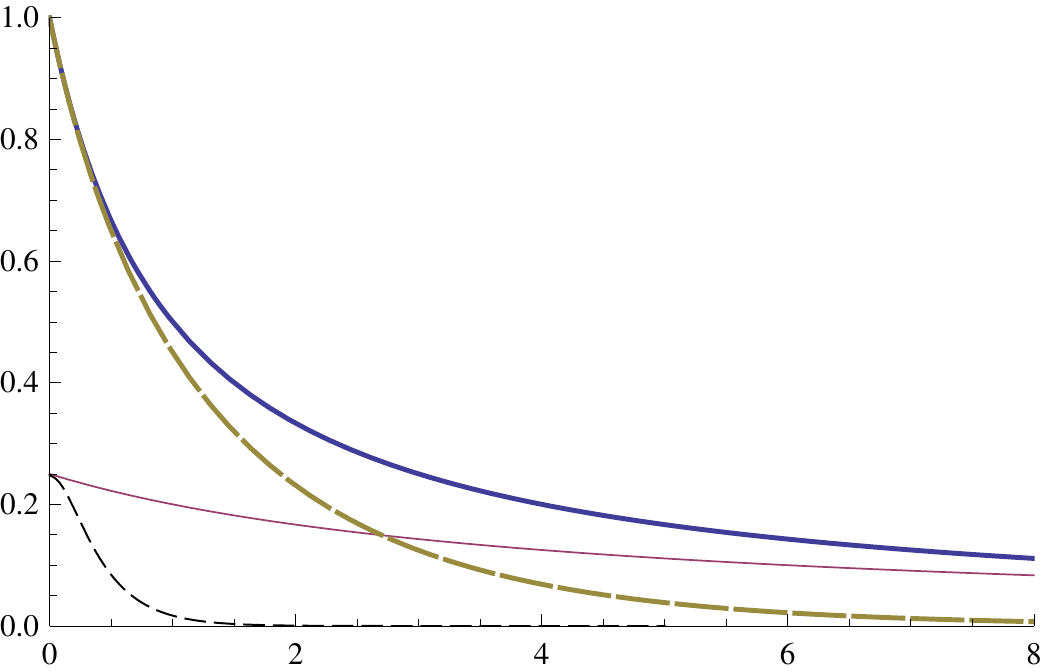}
  \put(-210,120){$\Prob(X=-4.95)=1-\Prob(X=0.05)=0.01$}
  \put(-180,100){$\Prob(X=1)=\Prob(X=-1)=0.5$}
  \put(-120,80){$X\sim \mathcal{N}(0,0.25)$}
  \put(-90,60){$X\sim \mathcal{N}(0,1)$}
  \put(-85,55){\vector(0,-1){20}}
  \put(-115,75){\vector(0,-1){44}}
  \put(-175,95){\vector(0,-1){53}}
  \put(-205,115){\vector(0,-1){100}}
  \put(-120,-5){$\snr$}
  \put(-258,170){$\mmse(X,\snr)$}
  }
  \caption{The MMSE of Gaussian and binary inputs as a function of the
    SNR.}
  \label{f:mmse}
\end{figure}
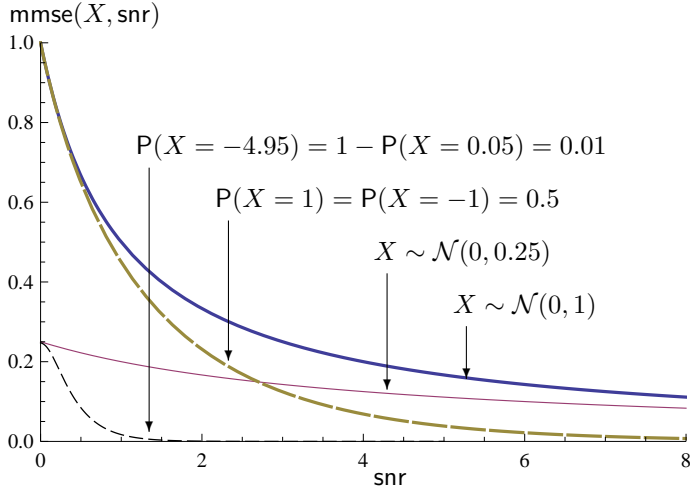%
%
The function $\mmse(X,\snr)$ is illustrated in Fig.~\ref{f:mmse} for
four special inputs: the standard Gaussian variable, a Gaussian
variable with variance $1/4$, as well as symmetric and asymmetric
binary random variables, all of zero mean.

Optimal estimation intrinsically underlies many fundamental
information theoretic results, which describe the boundary between
what is achievable and what is not, given unlimited computational
power.  
Simple quantitative connections between the MMSE and information
measures were revealed in~\cite{GuoSha05IT}.
%
One such result is that, for arbitrary but fixed $P_X$, 
\begin{equation} \label{eq:ie}
  \mmse(X,\snr) = 2 \pd{\snr} I(X;\sqrt{\snr}\,X+N).
\end{equation}
This relationship implies the following integral expression for the
mutual information: 
\begin{equation}  \label{eq:iei}
  I(X;\sqrt{\snr}\,g(X)+N) = \frac12 \int_0^\snr 
  \mmse( g(X), \gamma ) \diff \gamma
\end{equation}
which holds for any one-to-one real-valued function $g$.
%
By sending $\snr\rightarrow \infty$ in~\eqref{eq:iei}, we find the
entropy of every discrete random variable $X$ can be expressed as
(see~\cite{GuoSha05IT, VerGuo06IT}):
\begin{equation}  \label{eq:H}
  H(X) = \frac12 \int_0^\infty \mmse( g(X), \gamma ) \diff \gamma
\end{equation}
whereas the differential entropy of any continuous random variable $X$
can be expressed as:
\begin{equation}       \label{eq:hmv}
  h(X) = \frac{ \log\left( 2 \pi e \right) }2 -
  \frac12 \int_0^\infty  \frac1{1 + \gamma} - \mmse(g(X), \gamma) \diff \gamma.
\end{equation}

The preceding information--estimation relationships have found a
number of applications, e.g., in nonlinear filtering~\cite{GuoSha05IT,
  Weissm10IT}, in multiuser detection~\cite{GuoVer05IT}, in power
allocation over parallel Gaussian channels~\cite{LozTul06IT,
  PerRod10IT}, in the proof of Shannon's entropy power inequality
(EPI) and its generalizations~\cite{VerGuo06IT, GuoSha06ISIT,
  TulVer06IT}, and in the treatment of the capacity region of several
multiuser channels~\cite{TseYat09X, ZhuGuo11IT, BusLiu09EURASIP}.
Relationships between relative entropy and mean-square error are also
found in~\cite{Verdu10IT, Guo09ISIT}.  Moreover, many such results
have been generalized to vector-valued inputs and multiple-input
multiple-output (MIMO) models~\cite{GuoSha05IT, GuoSha06ISIT,
  PalVer06IT}.

Partially motivated by the important role played by the MMSE in
information theory, this paper presents a detailed study of the key
mathematical properties of $\mmse(X,\snr)$.  The remainder of the
paper is organized as follows.

In \prettyref{s:bnd}, we establish bounds on the MMSE as well as on
the conditional and unconditional moments of the conditional mean
estimation error.  In particular, it is shown that the tail of the
posterior distribution of the input given the observation vanishes at
least as quickly as that of some Gaussian density.  Simple properties
of input shift and scaling are also shown.

In \prettyref{s:analytic}, $\mmse(X,\snr)$ is shown to be an
infinitely differentiable function of $\snr$ on $(0,\infty)$ for every
input distribution regardless of the existence of its moments (even
the mean and variance of the input can be infinite).  Furthermore,
under certain conditions, the MMSE is found to be real analytic
at all positive SNRs, and hence can be arbitrarily well-approximated
by its Taylor series expansion. %

In \prettyref{s:dmmse}, the first three derivatives of the MMSE with
respect to the SNR are expressed in terms of the
average central moments of the input conditioned on the output.  The
result is then extended to the conditional MMSE.

\prettyref{s:cross} shows that the MMSE is concave in the
distribution $P_X$ at any given SNR.  The monotonicity of the MMSE of
a partial sum of independent identically distributed (i.i.d.) random
variables is also investigated.
It is well-known that the MMSE of a non-Gaussian input is dominated by
the MMSE of a Gaussian input of the same variance.  It is further
shown in this paper that the MMSE curve of a non-Gaussian input and
that of a Gaussian input cross each other at most once over
$\snr\in(0,\infty)$, regardless of their variances.


In \prettyref{s:app}, properties of the MMSE are used to establish
Shannon's EPI in the special case where one of the variables is Gaussian.
Sidestepping the EPI, the properties of the MMSE lead to simple and
natural proofs of the fact that Gaussian input is optimal for both the
Gaussian wiretap channel and the scalar Gaussian broadcast channel.


\section{Basic Properties}
\label{s:bnd}

\subsection{The MMSE}

The input $X$ and the observation $Y$ in the model described by
$Y=\sqrt{\snr}\,X+N$ are tied probabilistically by the conditional
Gaussian probability density function:
\begin{equation} \label{eq:pyx} %
  p_{Y|X}(y|x;\snr) = \varphi\left( y-\sqrt{\snr}\,x \right)
\end{equation}
where $\varphi$ stands for the standard Gaussian density:
\begin{equation}
  \varphi(t) = \frac1{\sqrt{2\pi}} e^{-\frac{t^2}2} \ .
\end{equation}
Let us define for every $a\in\reals$ and $i=0,1,\dots$,
\begin{equation}  \label{eq:hi}
  h_i(y; a) = \expect{X^i \varphi(y - a X)}
\end{equation}
which is always well defined because
$\varphi(y-ax)$ is bounded and vanishes quadratic exponentially fast
as either $x$ or $y$ becomes large with the other variable bounded.
In particular, 
$h_0(y;\sqrt{\snr})$
is nothing but the marginal distribution of the observation $Y$,
which is always strictly positive.
The conditional mean estimate
can be expressed as~\cite{GuoVer05IT, GuoSha05IT}:
\begin{equation}  \label{eq:exy}
\expcnd{X}{Y=y}
  = \frac{ h_1(y;\sqrt{\snr}) }{ h_0(y;\sqrt{\snr}) }
\end{equation}
and the MMSE can be calculated as~\cite{GuoVer05IT}:
\begin{equation} \label{eq:eq} %
  \begin{split}
    &\mmse(X,\snr) =\\
    &\iint_\reals
    \left( x -
      \frac{h_1(y;\sqrt{\snr})}{h_0(y;\sqrt{\snr})} \right)^2
    \varphi(y-\sqrt{\snr}\,x) \diff y\, \diff P_X(x)
  \end{split}
\end{equation}
which can be simplified if
$\expect{X^2}<\infty$: 
\begin{equation}  \label{eq:eq2}
  \mmse(X,\snr) = \expect{X^2} - \int_{-\infty}^\infty
  \frac{h_1^2(y;\sqrt{\snr})}{h_0(y;\sqrt{\snr})} \, \diff y.
\end{equation}

Note that the estimation error $X-\expcnd{X}{Y}$ remains the same if
$X$ is subject to a constant shift.  Hence the following well-known
fact:
\begin{proposition} \label{pr:shift}
  For every random variable $X$ and $a\in \reals$,
  \begin{align}  \label{eq:shift}
    \mmse( X + a, \snr ) = \mmse( X, \snr ).
  \end{align}
\end{proposition}

The following is also straightforward from the definition of MMSE.

\begin{proposition} \label{pr:sc}
  For every random variable $X$ and $a\in \reals$, 
  \begin{align}  \label{eq:sc}
    \mmse( a X, \snr ) = a^2 \, \mmse( X, a^2 \,\snr ). 
  \end{align}
\end{proposition}

\subsection{The Conditional MMSE and SNR Increment}

For any pair of jointly distributed variables $(X,U)$, the conditional
MMSE of estimating $X$ at SNR $\gamma\geq0$ given $U$ is defined as:
\begin{equation}  \label{eq:mmseu}
  \mmse(X,\gamma|U) = \Exp\big\{ \left(
    X - \expect{X|\sqrt{\gamma}\,X+N,U} \right)^2 \big\}
\end{equation}
where $N\sim \mathcal{N} (0,1)$ is independent of $(X,U)$.  It can be
regarded as the MMSE achieved with side information $U$ available to
the estimator.  
For every $u$, let $X_u$ denote a random variable indexed by $u$ with
distribution $P_{X|U=u}$.  Then the conditional MMSE can be seen as an
average:
\begin{equation}  \label{eq:exu} %
  \mmse(X,\snr|U) = \int \mmse(X_u,\snr) P_U(\diff u).
\end{equation}

A special type of conditional MMSE is obtained when the side
information is itself a noisy observation of $X$ through an
independent additive Gaussian noise channel.  It has long been noticed
that two independent looks through Gaussian channels is equivalent to
a single look at the sum SNR, e.g., in the context of maximum-ratio
combining.  As far as the MMSE is concerned, the SNRs of the direct
observation and the side information simply add up.

\begin{proposition}\label{pr:ic}
  For every $X$ and every $\snr, \gamma \ge0$,
  \begin{equation} \label{eq:ic}
    \mmse(X,\gamma|\sqrt{\snr}\,X+N) = \mmse(X,\snr+\gamma) 
  \end{equation}
  where $N\sim\mathcal{N} (0,1)$ is independent of $X$.
\end{proposition}

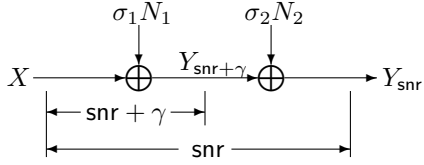
\begin{figure}
  \begin{center}
    \begin{picture}(150,60)(0,0)
      {\small
        \put(0,27){$X$}
        \put(65,33){$Y_{\snr+\gamma}$}
        \put(142,27){$Y_\snr$}
        \put(30,14){$\snr+\gamma$}
        \put(70,0){$\snr$}
        \put(40,52){$\sigma_1 N_1$}
        \put(90,52){$\sigma_2 N_2$}
        \put(15,0){\line(0,1){25}}
        \put(75,12){\line(0,1){13}}
        \put(130,0){\line(0,1){25}}
        \put(50,50){\vector(0,-1){15}}
        \put(100,50){\vector(0,-1){15}}
        \put(10,30){\vector(1,0){35}}
        \put(55,30){\vector(1,0){40}}
        \put(105,30){\vector(1,0){35}}
        \put(50,30){\makebox(0,0){$\bigoplus$}}
        \put(100,30){\makebox(0,0){$\bigoplus$}}
        \put(28,17){\vector(-1,0){13}}
        \put(62,17){\vector(1,0){13}}
        \put(65,3){\vector(-1,0){50}}
        \put(85,3){\vector(1,0){45}}
      }
    \end{picture}
  \end{center}
  \caption{An incremental Gaussian channel.}
  \label{f:ig}
\end{figure}

\prettyref{pr:ic} enables translation of the MMSE at any given
SNR to a conditional MMSE at a smaller SNR.  This result was first
shown in~\cite{GuoSha05IT} using the {\em incremental channel}\,
technique, and has been instrumental in the proof of
information--estimation relationships such as~\eqref{eq:ie}.
\prettyref{pr:ic} is also the key to the regularity properties
and the derivatives of the MMSE presented in subsequent sections.  A
brief proof of the result is included here for completeness.

\begin{IEEEproof}[Proof of \prettyref{pr:ic}]
Consider a cascade of two Gaussian channels as depicted in
Fig.~\ref{f:ig}:
\begin{subequations}  \label{eq:y12}%
  \begin{align}
    Y_{\snr+\gamma} &= X + \sigma_1 N_1 \label{eq:y1} \\
    Y_\snr &= Y_{\snr+\gamma} + \sigma_2 N_2 \label{eq:y2}
  \end{align}
\end{subequations}
where $X$ is the input, $N_1$ and $N_2$ are independent standard
Gaussian random variables.  A subscript is used to explicitly denote
the SNR at which each observation is made.  Let $\sigma_1, \sigma_2
>0$ satisfy $\sigma_1^2 = 1/(\snr + \gamma)$ and $\sigma_1^2 +
\sigma_2^2 = 1/\snr$ so that the SNR of the first channel
\eqref{eq:y1} is $\snr+\gamma$ and that of the composite channel is
$\snr$.  A linear combination of~\eqref{eq:y1} and~\eqref{eq:y2}
yields
\begin{equation}
  (\snr+\gamma)\, Y_{\snr+\gamma}
  = \snr \, Y_\snr + \gamma \, X + \sqrt{\gamma} \, W
  \label{eq:yyxw}
\end{equation}
where we have defined $W = ( \gamma\, \sigma_1\, N_1 - \snr\,
\sigma_2\, N_2 ) / \sqrt{\gamma}\,$.  Clearly, the input--output
relationship defined by the incremental channel~\eqref{eq:y12} is
equivalently described by~\eqref{eq:yyxw} paired with~\eqref{eq:y2}.
Due to mutual independence of $(X,N_1,N_2)$, it is easy to see that
$W$ is standard Gaussian and $(X,W,\sigma_1N_1 +\sigma_2N_2)$ are
mutually independent.  Thus $W$ is independent of $(X,Y_\snr)$
by~\eqref{eq:y12}.  Based on the above observations, the relationship
of $X$ and $Y_{\snr+\gamma}$ conditioned on $Y_\snr=y$ is exactly the
input--output relationship of a Gaussian channel with SNR equal to
$\gamma$ described by~\eqref{eq:yyxw} with $Y_\snr=y$.
Because $Y_\snr$ is a physical degradation of $Y_{\snr+\gamma}$,
providing $Y_\snr$ as the side information does not change the overall
MMSE, that is,
 $ \mmse(X|Y_{\snr+\gamma}) = \mmse( X, \gamma | Y_\snr )$,
which proves~\eqref{eq:ic}.
\end{IEEEproof}

\subsection{Bounds}

The input to a Gaussian model with nonzero SNR can always be estimated
with finite mean-square error based on the output, regardless of the
input distribution.  In fact, $\tilde{X} = Y/\sqrt{\snr}$ achieves
mean-square error of $1/\snr$, even if $\expect{X}$ does not exist.
Moreover, the trivial zero estimate achieves mean-square error of
$\expect{X^2}$. 

\begin{proposition} \label{pr:ub}
  For every input $X$,
  \begin{equation}  \label{eq:ub}
    \mmse(X,\snr) \leq \frac1{\snr}
  \end{equation}
  and in case the input variance $\var{X}$ is finite,
  \begin{equation}  \label{eq:ub2}
    \mmse(X,\snr) \leq \min \left\{ \var{X}, \frac1{\snr} \right\}.
  \end{equation}
\end{proposition}

\prettyref{pr:ub} can also be established using the fact that
$\snr\cdot\mmse(X,\snr) = \mmse(N|\sqrt{\snr}\,X+N) \le 1$, which is
simply because the estimation error of the input is proportional to
the estimation error of the noise~\cite{GuoSha06ISIT}:
\begin{align}  \label{eq:XN}
  \sqrt{\snr}(X - \expcnd{X}{Y})
  &=  
  \expcnd{N}{Y} - N \ .
\end{align}

Using~\eqref{eq:XN} and known moments of the Gaussian density, higher
moments of the estimation errors can also be bounded as shown in
\prettyref{a:mmt}:

\begin{proposition} \label{pr:mmt} %
  For every random variable $X$ and $\snr>0$,
  \begin{equation} \label{eq:mmt} %
    \expect{ \left| X - \expcnd{X}{\sqrt{\snr}\,X+N} \right|^n } \le
    \left( \frac2{\sqrt{\snr}} \right)^n \sqrt{n!}
  \end{equation}
  for every $n=0,1,\dots$, where $N\sim \mathcal{N} (0,1)$ is
  independent of $X$.
\end{proposition}

In order to show some useful characteristics of the posterior input
distribution, it is instructive to introduce the notion of {\em
  sub-Gaussianity}.  A random variable $X$ is called sub-Gaussian if
the
tail of its distribution is dominated by that of some Gaussian random
variable, i.e.,
\begin{equation}	\label{eq:subg.cdf}
  \Prob(|X| > \lambda) \le C e^{ -c \lambda^2}
\end{equation}
for some $c, C>0$ and all $\lambda > 0$.  Sub-Gaussianity can be
equivalently characterized by that the growth of moments or moment
generating functions does not exceed those of some
Gaussian~\cite[Theorem 2]{subgaussian}.
 
\begin{lemma}
The following statements are equivalent:
\begin{enumerate}
\item $X$ is sub-Gaussian;
\item There exists $C > 0$ such that for every $k=1,2,\dots$,
	\begin{equation}
\expect{|X|^{k}} \leq C^k \sqrt{k!} \ ;
	\label{eq:subg.moments}
\end{equation}	
\item There exist $c, C > 0$ such that for all $t > 0$,
	\begin{equation}
	\expect{e^{t X}} \leq C e^{c\, t^2} \ .
	\label{eq:subg.MGF}
\end{equation}
\end{enumerate}
\label{lmm:subg.char}		
\end{lemma}

Regardless of the prior input distribution, the posterior distribution
of the input given the noisy observation through a Gaussian channel is
always sub-Gaussian, and the 
posterior moments can be upper bounded.  This is formalized in the
following result
proved in~\prettyref{a:subg.post}:

\begin{proposition}
  Let $X_y$ be distributed according to $P_{X|Y = y}$ where $Y = a X +
  N$, $N\sim\mathcal{N}(0,1)$ is independent of $X$, and $a \neq 0$.
  Then $X_y$ is sub-Gaussian for every $y \in \reals$. Moreover,
  \begin{equation} \label{eq:Xy.Q} %
    \Probk{|X_y| \geq x} \leq \sqrt{\frac2{\pi}}
    \frac{e^{\frac{y^2}2}}{ h_0(y; a)} e^{ -\frac{a^2x^2}{4} }
  \end{equation}
  and, for every $n=1,2,\dots$,
  \begin{equation} \label{eq:Xyn}
    \expect{ |X_y|^n } \leq \frac{n e^{\frac{y^2}{2}}
    }{h_0(y; a)} \left(\frac{\sqrt{2}}{|a|} \right)^n \sqrt{(n-1)!} 
  \end{equation}
  and
  \begin{equation}  \label{eq:mmty}
    \expect{ \left| X_y - \expect{X_y} \right|^n }
    \le 2^n \, \expect{ |X_y|^n } \ .
  \end{equation}
	\label{pr:subg.post}
\end{proposition}

\section{Smoothness and Analyticity}
\label{s:analytic}

This section studies the regularity of the MMSE as a function of the
SNR, where the input distribution is arbitrary but fixed.  In
particular, it is shown that $\mmse(X,\snr)$ is a smooth function of
$\snr$ on $(0, \infty)$ for every $P_X$.  This conclusion clears the
way towards calculating its derivatives in \prettyref{s:dmmse}.
Under certain technical conditions, the MMSE is also found to be real
analytic in $\snr$.
  This implies that the MMSE can be
reconstructed from its local derivatives.
As we shall see, the regularity of the MMSE at the point of zero SNR
requires additional conditions.

\subsection{Smoothness}

\begin{proposition}
  For every $X$, $\mmse(X,\snr)$ is infinitely differentiable at every
  $\snr > 0$.  If $\expect{X^{k+1}}<\infty$, then $\mmse(X,\snr)$ is
  $k$ right-differentiable at $\snr = 0$.  Consequently, $\mmse(X,
  \snr)$ is infinitely right differentiable at $\snr = 0$ if all
  moments of $X$ are finite.
	\label{pr:mmse.smooth}
\end{proposition}

\begin{IEEEproof}
  The proof is divided into two parts.  In the first part we first
  establish the smoothness assuming that all input moments are finite,
  i.e., $\expect{X^k}<\infty$ for all $k=1,2,\dots$.


  For convenience, let $Y=aX+N$ where $a^2=\snr$.  For every
  $i=0,1,\dots$, denote
  \begin{equation}
    g_i(y;a)=\fracpk{}{a}{i}\left(\frac{h_1^2}{h_0}\right)(y; a) \label{eq:gi}
  \end{equation}
  and
  \begin{equation}
    m_i(a) = \int_{-\infty}^\infty g_i(y; a) \diff y \label{eq:mi}
  \end{equation}
  where $h_i$ is given by~\eqref{eq:hi}.  By~\eqref{eq:eq2}, we have
  \begin{equation}
    \mmse(X,a^2) = 
    \expect{X^2} - m_0(a).
    \label{eq:m0}
  \end{equation}
  We denote by $H_n$ the $n$-th Hermite polynomial~\cite[Section
  5.5]{orthogonal.poly}:
\begin{align}
  H_n(x)
  &= \frac{(-1)^n}{\varphi(x)} \fracdk{\varphi(x)}{x}{n}  \label{eq:Hn} \\
  &= n! \sum_{k = 0}^{\lfloor \frac{n}{2}\rfloor}
  \frac{(-1)^k}{k!(n-2k)!}  (2x)^{n-2k} \ . \label{eq:Hn.2}
\end{align}
Denote $h^{(n)}_i(y;a) = \partial^n h_i(y;a)/\partial a^n$ throughout
the paper.  Then 
\begin{align}
  \frac{h^{(n)}_i(y;a)}{h_0(y;a)}
  &= \frac1{h_0(y;a)} \expect{X^{i+n} H_n(y - a X) \varphi(y - a X)}
  \label{eq:hi.1} \\
  &= \expect{X^{i+n} H_n(N) \mid Y = y}	\label{eq:hi.exp}
\end{align}
where the derivative and expectation can be exchanged to
obtain~\eqref{eq:hi.1} because the product of any polynomial and the
Gaussian density is bounded.

The following lemma is established in \prettyref{a:gi}:

\begin{lemma} \label{lmm:gi}
  For every $i=0,1,\dots$ and all $w > v$, $(y, a) \mapsto g_i(y; a)$
  is integrable on $\reals \times [v, w]$.
\end{lemma}

Using \prettyref{lmm:gi} and~\eqref{eq:mi}, we have
\begin{align}
  \int_{v}^w m_{i+1}(a) \diff a &= \int_v^w \int_{-\infty}^\infty
  g_{i+1}(y; a) \diff y \diff a \label{eq:fub.1}\\
  &= \int_{-\infty}^\infty g_i(y;w) - g_i(y;v) \diff y 	\label{eq:fub.3} \\
  &= m_{i}(w) - m_{i}(v) \label{eq:fub.4}
\end{align}
where~\eqref{eq:fub.3} is due to~\eqref{eq:gi} and Fubini's theorem.
Therefore for every $i \geq 0$, $m_i$ is continuous. Hence for each $a
\in \reals$,
\begin{equation}
	\fracd{m_{i}(a)}{a} = m_{i+1}(a)
	\label{eq:mi1}
\end{equation}
follows from the fundamental theorem of calculus~\cite[p. 97]{Royden88}.
In view of~\eqref{eq:m0}, we have
\begin{equation}
	\fracdk{\mmse(X,a^2)}{a}{i} = - m_i(a).
	\label{eq:dim}
\end{equation}
This proves that $a \mapsto \mmse(X, a^2) \in C^{\infty}(\reals)$,
which implies that $\mmse(X, \snr)$ is infinitely differentiable in
$\snr$ on $(0,\infty)$.

In the second part of this proof, we eliminate the requirement that
all moments of the input exist by resorting to the incremental-SNR
result, \prettyref{pr:ic}.  Fix arbitrary $\gamma>0$ and let $Y_\gamma
= \sqrt{\gamma}\, X + N$.  For every $u\in\reals$, let $X_{u;\gamma}
\sim P_{X|Y_\gamma=u}$.  By~\eqref{eq:eq2},~\eqref{eq:exu} and
\prettyref{pr:ic}, we have
\begin{align}
  \mmse(X, \gamma+a^2) 
  &= \int  \mmse(X_{u;\gamma}, a^2) P_{Y_{\gamma}}(\diff u) \\
  &= 
  \expect{X^2} - \tilde{m}_0(a)
  \label{eq:incremental}
\end{align}	
where
\begin{equation}
  h_i(y; a|u; \gamma) = \expcnd{X^i \varphi(y - a X)}{Y_\gamma=u}
  \label{eq:hibar}
\end{equation}
\begin{align}
  g_i(y;a|u;\gamma) = \fracpk{}{a}{i}\left(
    \frac{h_1^2}{h_0}\right)(y;a|u;\gamma) \label{eq:giu}
\end{align}
and
\begin{align}
  \tilde{m}_i(a) = \int_\reals \int_\reals g_i(y; a|u;\gamma) \diff y
  \, h_0(u;\gamma) \diff u
  \label{eq:miu}
\end{align}
for $i=0,1,\dots$.
By \prettyref{pr:mmt}, for each $u$, all moments of $X_{u;\gamma}$ are
finite.  Each $\tilde{m}_i$ is a well-defined real-valued function 
on $\reals$.
Repeating the first part of this proof with $h_i(y; a)$ replaced by
$h_i(y; a | u;\gamma)$, we conclude that $a \mapsto \mmse(X,
\gamma+a^2) \in C^{\infty}$ in $a$ at least on $|a|\ge\sqrt{\gamma}$,
which further implies that $a \mapsto \mmse(X, a^2) \in
C^{\infty}(\reals \backslash [-\sqrt{2\gamma},\sqrt{2\gamma}])$
because $a \mapsto \sqrt{a^2 -\gamma}$ has bounded derivatives of all
order when $|a| > \sqrt{2\gamma}$.  By the arbitrariness of $\gamma$,
we have $a \mapsto \mmse(X, a^2) \in C^{\infty}(\reals \backslash
\{0\})$, hence $\mmse(X, \cdot) \in C^{\infty}((0,\infty))$.

Finally, we address the case of zero SNR.  It follows
from~\eqref{eq:hi.exp} and the independence of $X$ and $Y$ at zero SNR
that
\begin{equation}
  \frac{1}{h_0} \fracpk{h_i}{a}{n}(y; 0) = \expect{X^{i+n}}	H_n(y).
	\label{eq:hi.exp.0}
\end{equation}
Since $\expect{|H_n(N)|} \le \sqrt{\expect{H_n^2(N)}} = \sqrt{n!}$ is always
finite, induction reveals that the $n$-th derivative of $m_0$ at $0$
depends on the first $n+1$ moments of $X$.
By Taylor's theorem and the fact that $m_0(a)$ is an even function of $a$,
we have
\begin{equation}
  m_0(a) = \sum_{j=0}^{i} \frac{m_{2j}(0)}{(2j)!} a^{2j} + \Bigo{|a|^{2i+2}}
\end{equation}
in the vicinity of $a=0$, which implies that $m_0$ is $i$
differentiable with respect to $a^2$ at $0$, with $\diff^i
m_0(0+)/\diff(a^2)^i
= m_{2i}(0)$, as long as $\expect{X^{i+1}}<\infty$.
\end{IEEEproof}

\subsection{Real Analyticity}
\label{s:ana}

A function $f: \reals \to \reals$ is said to be {\em real analytic} at
$x_0$ if it can be represented by a convergence power series in some
neighborhood of $x_0$, i.e., there exists $\delta>0$ such that
\begin{equation}
	f(x) = \sum_{n=0}^{\infty} a_n (x-x_0)^n
	\label{eq:series}
\end{equation}
for every $x\in(x_0-\delta,x_0+\delta)$.
One necessary and sufficient condition for $f$ to be real analytic is
that $f$ can be extended to some open disk $D(x_0, \delta) \triangleq
\{z \in \complex: |z-x_0| < \delta\}$ in the complex plane by the
power series~\eqref{eq:series}~\cite{Karuna05}.

\begin{proposition} \label{pr:mmse.ana}
  As a function of $a$, $\mmse(X, a^2)$ is real analytic at
  $a_0\in\reals$ if either one of the following two sets of conditions
  holds:
  \begin{enumerate}
  \item $X$ is sub-Gaussian, and there exist $c > 0$ and $r > 0$ such
    that for every $y\in\reals$,
    \begin{equation}
      \inf_{z \in D(a_0,r)} |h_0(y; z)| > 0
      \label{eq:h0.nonz}
    \end{equation}
    and
    \begin{equation}
      \liminf_{|y| \to \infty} \inf_{z \in D(a_0,r)} \frac{
        |h_0(y; z)|}{h_0(y; \Rerm{z})} > c
      \label{eq:h0lb}
    \end{equation}
  \item
    $a_0 \neq 0$, and there exist $c > 0$, $r>0$ and $\delta \in (0,a_0^2)$
    such that for every $y, u\in\reals$,
    \begin{equation}
      \inf_{z \in D(a_0,r)} |h_0(y; z|u, \delta)| > 0
      \label{eq:h0.nonz.in}
    \end{equation}
    and
    \begin{equation}
      \liminf_{|y| \to \infty} \inf_{z \in D(a_0,r)} \frac{
        |h_0(y; z|u, \delta)|}{h_0(y; \Rerm{z}|u, \delta)} > c \ .
      \label{eq:h0lb.in}
    \end{equation}
  \end{enumerate}
  Moreover, whenever $\mmse(X, a^2)$ is real analytic at $a\in\reals$,
  the function $\mmse(X,\snr)$ is also analytic at $\snr=a^2$.
\end{proposition}

The last statement in \prettyref{pr:mmse.ana} is because of the
following.  The Taylor series expansion of $\mmse(X,a^2)$ at $a=0$ is
an even function, so that the analyticity of $\mmse(X,a^2)$ at $a=0$
implies the anlyticity of $\mmse(X,\snr)$ at $\snr=0$.  If
$\mmse(X,a^2)$ is analytic at $a\ne0$, then $\mmse(X,\snr)$ is also
analytic at $\snr=a^2$ because $\snr\mapsto \sqrt{\snr}$ is real
analytic at $\snr>0$, and composition of analytic functions is
analytic~\cite{KraPar02}.  It remains to establish the analyticity of
$a\mapsto \mmse(X,a^2)$, which is relegated to \prettyref{a:ana}.

Conditions~\eqref{eq:h0.nonz} and~\eqref{eq:h0lb} can be
understood as follows. Recall that $h_0(y; a)
$ denotes the density of $Y = a X + N$.  The function $h_0(y;a)$ stays
positive for all $a\in\reals$, and 
decays no faster than the Gaussian density.  However, $h_0(y;a)$ may
vanish for some $a\in\complex$, so that the MMSE may not be extendable
to the convex plane.  Hence the purpose of \prettyref{eq:h0.nonz}
and~\eqref{eq:h0lb} is to ensure that the imaginary part of $a$ has
limited impact on $|h_0|$.

As an example, consider the case where $X$ is equiprobable on $\{\pm
1\}$.  Then 
\begin{equation}
  h_0(y; a) = \varphi(y) \exp(-a^2/2) \cosh(ay)\ .  
\end{equation}
Letting $a = j t$ yields $h_0(y; j t) = \varphi\left(\sqrt{y^2 -
    t^2}\right) \cos(ty)$, which has infinitely many zeros. In fact,
in this case the MMSE is given by~\eqref{eq:mmseb}, or in an
equivalent form:
\begin{equation} \label{eq:mmse.binary} %
  \mmse(X, a^2) = 1 - \int^\infty_{-\infty} \varphi(y) \tanh\left( a^2
    - a y \right) \diff y.
\end{equation}  
Then for any $r > 0$, there exists $|a_0| < r$ and $y_0 \in \reals$,
such that $a_0^2 - a_0 y_0 = j \frac{\pi}{2}$ and the integral in
\prettyref{eq:mmse.binary} diverges near $y_0$. Therefore $\mmse(X,
a^2)$ cannot be extended to any point on the imaginary axis, hence it
is not real analytic at $a = 0$. Nevertheless, when $\Rerm{a} \neq 0$,
condition~\eqref{eq:h0lb} is satisfied. Hence $\mmse(X, a^2)$ is real
analytic on the real line except zero, which can be shown from
\prettyref{eq:mmse.binary} directly. Similarly, for any
finite-alphabet, 
exponential or Gaussian distributed $X$,
\prettyref{eq:h0.nonz.in} and~\eqref{eq:h0lb.in} can be verified for
all $a \neq 0$, hence the corresponding MMSE is real analytic at all
positive SNR.

\section{Derivatives}
\label{s:dmmse}

\subsection{Derivatives of the MMSE}
\label{s:dMi}

With the smoothness of the MMSE established in
\prettyref{pr:mmse.smooth}, its first few derivatives with respect to
the SNR are explicitly calculated in this section.  Consider first the
Taylor series expansion of the MMSE around $\snr=0^+$ to the third
order:\footnote{The previous result for the expansion of $\mmse(\snr)$
  around $\snr=0^+$, given by equation (91) in~\cite{GuoSha05IT} is
  mistaken in the coefficient corresponding to $\snr^2$.  The
  expansion of the mutual information given by (92)
  in~\cite{GuoSha05IT} should also be corrected accordingly.  The
  second derivative of the MMSE is mistaken in~\cite{GuoSha08ISIT} and
  corrected in \prettyref{pr:ds} in this paper.  The function
  $\mmse(X,\snr)$ is not always convex in $\snr$ as claimed
  in~\cite{GuoSha08ISIT}, as illustrated using an example in
  Fig.~\ref{f:mmse}.}
\begin{equation}  \label{eq:taylor}
  \begin{split}
    &\mmse(X,\snr) = 1 - \snr + \left[ 2-(\Exp
      X^3)^2 \right]\frac{\snr^2}{2}  \\
    & -\!\left[ 15 - 12(\Exp X^3)^2\! - 6\Exp X^4 + (\Exp X^4)^2
    \right]\!  \frac{\snr^3}{6}\! + \mathcal{O}(\snr^4)
  \end{split}
\end{equation}
where $X$ is assumed to have zero mean and unit variance.
The first three derivatives of the MMSE at $\snr=0^+$ are thus evident
from~\eqref{eq:taylor}.
The technique for obtaining~\eqref{eq:taylor}
is to expand~\eqref{eq:pyx} in terms of the small signal
$\sqrt{\snr}\,X$, evaluate $h_i(y;\sqrt{\snr})$ given by~\eqref{eq:hi}
at the vicinity of $\snr=0$ using the moments of $X$ (see equation
(90) in~\cite{GuoSha05IT}), and then calculate~\eqref{eq:eq}, where
the integral over $y$ can be evaluated as a Gaussian integral.


The preceding expansion of the MMSE at $\snr=0^+$ can be lifted to
arbitrary SNR using the SNR-incremental result, \prettyref{pr:ic}.
Finiteness of the input moments is not required for $\snr>0$ because
the conditional moments are always finite due to \prettyref{pr:mmt}.

For notational convenience, we define the following random variables:
\begin{equation} \label{eq:Mi} %
  M_i = \expcnd{ \left( X - \expcnd{X}{Y} \right)^i }{ Y }, \quad
  i=1,2,\dots
\end{equation}
which, according to \prettyref{pr:mmt}, are well-defined in case
$\snr>0$, and reduces to the unconditional moments of $X$ in case
$\snr=0$.  Evidently, $M_1=0$, $M_2 = \var{X|\sqrt{\snr}\,X+N}$ and
\begin{equation} \label{eq:em2}
  \expect{ M_2 } = \mmse( X, \snr ).
\end{equation}
If the input distribution $P_X$ is symmetric, 
then the distribution of $M_i$ is also symmetric for all odd $i$.

The derivatives of the MMSE are found to be the expected value of
polynomials of $M_i$, whose existence is guaranteed by
\prettyref{pr:mmt}.

\begin{proposition} \label{pr:ds}
  For every random variable $X$ and every $\snr>0$,
  \begin{align}  \label{eq:ds1}
    \frac{ \diff\, \mmse(X,\snr) }{\diff\, \snr}
    &= - \expect{ M_2^2 } 
  \end{align} 
  \begin{align}    \label{eq:ds2}
    \frac{ \diff^2 \mmse(X,\snr) }{\diff\, \snr^2}
    &= \expect{ 2 M_2^3 - M_3^2 } 
  \end{align}
  and
  \begin{equation}    \label{eq:ds3}
    \begin{split}
      & \frac{ \diff^3 \mmse(X,\snr) }{\diff\, \snr^3} 
      = \expect{ 6 M_4 M_2^2 - M_4^2 + 12 M_3^2 M_2 - 15 M_2^4 }.
    \end{split}
  \end{equation}
  The three derivatives are also valid at $\snr=0^+$ if $X$ has finite
  second, third and fourth moment, respectively.
\end{proposition}

We relegate the proof of \prettyref{pr:ds} to \prettyref{a:ds}.  It is
easy to check that the derivatives found in \prettyref{pr:ds} are
consistent with the Taylor series expansion~\eqref{eq:taylor} at zero
SNR.

In light of the proof of \prettyref{pr:mmse.smooth}
(and~\eqref{eq:dim}), the Taylor series expansion of the MMSE can be
carried out to arbitrary orders, so that all derivatives of the MMSE
can be obtained as the expectation of some polynomials of the
conditional moments, although the resulting expressions become
increasingly complicated.

\prettyref{pr:ds} is easily verified in the special case of standard
Gaussian input ($X\sim \mathcal{N} (0,1)$), where conditioned on
$Y=y$, the input is Gaussian distributed:
\begin{equation}
  X\sim \mathcal{N} \left( 
    \frac{\sqrt{\snr}} {1+\snr} y, \frac1{1+\snr} \right).
\end{equation}
In this case $M_2=(1+\snr)^{-1}$, $M_3=0$ and $M_4=3(1+\snr)^{-2}$ are
constants, and~\eqref{eq:ds1},~\eqref{eq:ds2} and~\eqref{eq:ds3} are
straightforward.

\subsection{Derivatives of the Mutual Information}

Based on \prettyref{pr:mmse.ana} and~\ref{pr:ds}, the following
derivatives of the mutual information are extensions of the key
information-estimation relationship~\eqref{eq:ie}.

\begin{corollary} \label{cr:di}
  For every distribution $P_X$ and $\snr>0$,
  \begin{equation}    \label{eq:di}
    \frac{\diff^i}{\diff \snr^i} I(X;\sqrt{\snr}\,X+N)
    = \frac{(-1)^{i-1}}{2} \,\expect{ M_2^i }
  \end{equation}
  for $i=1,2$, 
  \begin{equation}    \label{eq:di.3}
    \frac{\diff^3}{\diff \snr^3} I(X;\sqrt{\snr}\,X+N)
    = \expect{ M_2^3 - \frac12 M_3^2 }
  \end{equation}
and
  \begin{equation}
    \begin{split}
      \frac{\diff^4}{\diff \snr^4} & I(X;\sqrt{\snr}\,X+N) \\
      &= \frac12 \expect{ - M_4^2 + 6 M_4 M_2^2 + 2 M_3^2 M_2 - 15 M_2^4 }.
      \label{eq:i4}
    \end{split}
  \end{equation}
  as long as the corresponding expectation on the right hand side
  exists.  In case one of the two set of conditions in
  \prettyref{pr:mmse.ana} holds, $\sqrt{\snr}\mapsto
  I(\sqrt{\snr}\,X+N; X)$ is also real analytic.
\end{corollary}

\prettyref{cr:di} is a generalization of previous results on the
small SNR expansion of the mutual information such as
in~\cite{PreVer04IT}.  Note that~\eqref{eq:di} with $i=1$ is exactly
the original relationship of the mutual information and the MMSE given
by~\eqref{eq:ie} in light of~\eqref{eq:em2}.

\subsection{Derivatives of the Conditional MMSE}

The derivatives in \prettyref{pr:ds} can be generalized to the
conditional MMSE defined in~\eqref{eq:mmseu}.
The following is a straightforward extension of~\eqref{eq:ds1}.

\begin{corollary} \label{cr:ds} For every jointly distributed $(X,U)$
  and $\snr>0$,
  \begin{align}  \label{eq:dsu}
    \pd{\snr} \mmse(X, &\snr|U) = - \expect{ 
      M^2_2(U) 
    } 
  \end{align}
  where for every $u$ and $i=1,2,\dots$,
  \begin{equation} \label{eq:Miu}
    M_i(u) = 
    \expect{ \left[ X_u - \expect{X_u|Y} \right]^i \Big| Y
      = \sqrt{\snr}X_u+N }
  \end{equation}
  is a random variable dependent on $u$.
\end{corollary}

\section{Properties of the MMSE Functional}
\label{s:cross}

For any fixed $\snr$, $\mmse(X,\snr)$ can be regarded as a functional
of the input distribution $P_X$.  Meanwhile, the MMSE curve, $\{
\mmse(X,\snr), \snr\in [0,\infty)\}$, can be regarded as a
``transform'' of the input distribution.

\subsection{Concavity in Input Distribution}

\begin{proposition} \label{pr:ec}
  The functional $\mmse(X,\snr)$ is concave in $P_X$ for every
  $\snr\geq0$,
\end{proposition}

\begin{IEEEproof}
  Let $B$ be a Bernoulli variable with probability $\alpha$ to be 0.
  Consider any random variables $X_0$, $X_1$ independent of $B$.  Let
  $Z=X_B$, whose distribution is $\alpha P_{X_0} + (1-\alpha)
  P_{X_1}$.  Consider the problem of estimating $Z$ given $\sqrt{\snr}
  \,Z+N$ where $N$ is standard Gaussian.  Note that if $B$ is
  revealed, one can choose either the optimal estimator for $P_{X_0}$
  or $P_{X_1}$ depending on the value of $B$, so that the average MMSE
  can be improved.  Therefore,
  \begin{align} \label{eq:ea}
    \mmse( &Z, \,\snr ) \geq \mmse(Z,\snr|B) \\
    &= \alpha \mmse(X_0,\snr) + (1-\alpha) \mmse(X_1,\snr)
  \end{align}
  which proves the desired concavity.\footnote{Strict concavity is
    shown in~\cite{WuVer10ISIT}.}
\end{IEEEproof}


\subsection{Conditioning Reduces the MMSE}

As a fundamental measure of uncertainty, the MMSE decreases with
additional side information available to the estimator.  This is
because that an informed optimal estimator performs no worse than any
uninformed estimator by simply discarding the side information.

\begin{proposition}  \label{pr:proc}
  For any jointly distributed $(X,U)$ and $\snr\ge0$,
  \begin{equation}    \label{eq:proc}
    \mmse(X,\snr|U) \leq \mmse(X,\snr).
  \end{equation}
  For fixed $\snr>0$, the equality holds if and only if $X$ is
  independent of $U$.
\end{proposition}

\begin{IEEEproof}
  The inequality~\eqref{eq:proc} is straightforward by the concavity
  established in \prettyref{pr:ec}.  In case the equality holds,
  $P_{X|U=u}$ must be identical for $P_U$-almost every $u$ due to
  strict concavity~\cite{WuVer10ISIT}, that is, $X$ and $U$ are
  independent.
\end{IEEEproof}

\subsection{Monotonicity}

Propositions~\ref{pr:ec} and~\ref{pr:proc}
suggest that a mixture of random variables is
harder to estimate than the individual variables in average.  A
related result in~\cite{VerGuo06IT} states that a linear combination
of two random variables $X_1$ and $X_2$ is also harder to estimate
than the individual variables in some average:
\begin{proposition}[\!\!\cite{VerGuo06IT}]  \label{pr:cos}
  For every $\snr\geq0$ and $\alpha\in[0,2\pi]$,
  \begin{equation} \label{e:stm}
    \begin{split}
      &\mmse ( \cos\alpha\, X_1 + \sin\alpha\, X_2, \snr) \\
      &\quad\geq\cos^2\alpha \,\mmse(X_1,\snr) + \sin^2 \alpha \, \mmse(X_2,\snr)
    \end{split}
  \end{equation}
\end{proposition}

A generalization of \prettyref{pr:cos} concerns the MMSE of estimating
a normalized sum of
independent random variables.  Let $X_1,X_2,\dots$ be i.i.d.\ with finite
variance and $S_n=(X_1+\dots+X_n)/\sqrt{n}$.
It has been shown that the entropy of $S_n$ increases monotonically to
that of a Gaussian random variable of the same
variance~\cite{ArtBal04JAMS, TulVer06IT}.  
The following monotonicity result of the MMSE of estimating $S_n$ in
Gaussian noise can be established.

\begin{proposition} \label{pr:en}
  Let $X_1,X_2,\dots$ be i.i.d.\ with finite variance.  Let
  $S_n=(X_1+\dots+X_n)/\sqrt{n}$.  Then for every $\snr \geq 0$, 
  \begin{equation} \label{eq:esn}
    \mmse( S_{n+1}, \snr ) \geq \mmse( S_n, \snr ).
  \end{equation}
\end{proposition}

Because of the central limit theorem, as $n\rightarrow\infty$ the MMSE
converges to the MMSE of estimating a Gaussian random variable with
the same variance as that of $X$.

\prettyref{pr:en} is a simple corollary of the following general
result in~\cite{TulVer06IT}.

\begin{proposition}[\!\!{\cite{TulVer06IT}}]  \label{pr:tv}
  Let $X_1,\dots,X_n$ be independent.  For any $\lambda_1, \dots,
  \lambda_n\geq 0$ which sum up to one and any $\gamma\ge0$,
  \begin{equation} \label{eq:tv}
    \mmse\left( \sum^n_{i=1} X_i, \gamma \right) \geq
    \sum^n_{i=1} \lambda_i\, \mmse\left( \frac{ X_{\backslash i} }{
        \sqrt{(n-1) \lambda_i}}, \gamma \right)
  \end{equation}
  where $X_{\backslash i} = \sum\limits^n_{\substack{j=1,j\neq i}} X_j$\ .
\end{proposition}



Setting $\lambda_i=1/n$ in~\eqref{eq:tv} yields \prettyref{pr:en}.

In view of the representation of the entropy or differential entropy
using the MMSE in \prettyref{s:int}, integrating both sides
of~\eqref{eq:esn} proves a monotonicity result of the entropy or
differential entropy of $S_n$ whichever is well-defined.
More generally,~\cite{TulVer06IT} applies~\eqref{eq:hmv} and
\prettyref{pr:tv} to prove a more general result, originally given
in~\cite{ArtBal04JAMS}.

\subsection{Gaussian Inputs Are the Hardest to Estimate}

Any non-Gaussian input achieves strictly smaller MMSE than Gaussian
input of the same variance.  This well-known result is illustrated in
Fig.~\ref{f:mmse} and stated as follows.

\begin{proposition} \label{pr:ge}
  For every $\snr\geq0$ and random variable $X$ with variance no
  greater than $\sigma^2$,
  \begin{equation} \label{eq:ge}
    \mmse(X,\snr) \leq \frac{\sigma^2}{1+\snr\,\sigma^2}.
  \end{equation}
  The equality of~\eqref{eq:ge} is achieved if and only if the
  distribution of $X$ is Gaussian with variance $\sigma^2$.
\end{proposition}

\begin{IEEEproof}
  Due to Propositions \ref{pr:shift} and \ref{pr:sc}, it is enough to
  prove the result assuming that $\expect{X}=0$ and
  $\var{X}=\sigma_X^2$.
  Consider the linear estimator for the channel~\eqref{eq:yxn}:
  \begin{equation}
    \hat{X}^l = \frac{ \sqrt{\snr} }{ \snr\,\sigma_X^2 + 1 } Y
  \end{equation}
  which achieves the least mean-square error among all linear
  estimators, which is exactly the right hand side of~\eqref{eq:ge},
  regardless of the input distribution.  The inequality~\eqref{eq:ge}
  is evident due to the suboptimality of the linearity restriction on
  the estimator.
  The strict inequality is established as follows: If the linear
  estimator is optimal, then $ 
  \Exp\big\{ Y^k (X-\hat{X}^l) \big\} = 0 $ 
  for every $k=1,2,\dots$, due to the orthogonality principle.  It is
  not difficult to check that all moments of $X$ have to coincide with
  those of $\mathcal{N}(0,\sigma^2)$.  By Carleman's
  Theorem~\cite{Feller71}, the distribution is uniquely determined by
  the moments to be Gaussian.
\end{IEEEproof}

Note that in case the variance of $X$ is infinity,~\eqref{eq:ge}
reduces to~\eqref{eq:ub}.

\subsection{The Single-Crossing Property}

In view of \prettyref{pr:ge} and the scaling property of the
MMSE, at any given SNR, the MMSE of a non-Gaussian input is equal to
the MMSE of some Gaussian input with reduced variance.  The following
result suggests that there is some additional simple ordering of the
MMSEs due to Gaussian and non-Gaussian inputs.

\begin{proposition}[Single-crossing Property] \label{pr:uniq}
  For any given random variable $X$, the curve of $\mmse(X,\gamma)$ 
  crosses the curve of $(1+\gamma)^{-1}$, which is the MMSE function of the standard
  Gaussian distribution, at most once on $(0,\infty)$.
  Precisely, define
  \begin{equation} \label{eq:fg}
    f(\gamma) = (1+\gamma)^{-1} - \mmse(X,\gamma)
  \end{equation}
  on $[0,\infty)$.  Then
  \begin{enumerate}
  \item $f(\gamma)$ is strictly increasing at every $\gamma$ with $f(\gamma)<0$;
  \item   If $f(\snr_0)=0$, then $f(\gamma)\ge0$ at every $\gamma > \snr_0$;
  \item $\liminfty{\gamma} f(\gamma) = 0$.
  \end{enumerate}
  Furthermore, all three statements hold if the term $(1+\gamma)^{-1}$
  in~\eqref{eq:fg} is replaced by $\sigma^2/(1+\sigma^2\gamma)$ with
  any $\sigma$, which is the MMSE function of a Gaussian variable with
  variance $\sigma^2$.
\end{proposition}

\begin{figure}[htbp]
  \small
  \centering
  \includegraphics[width=\myfigwidth]{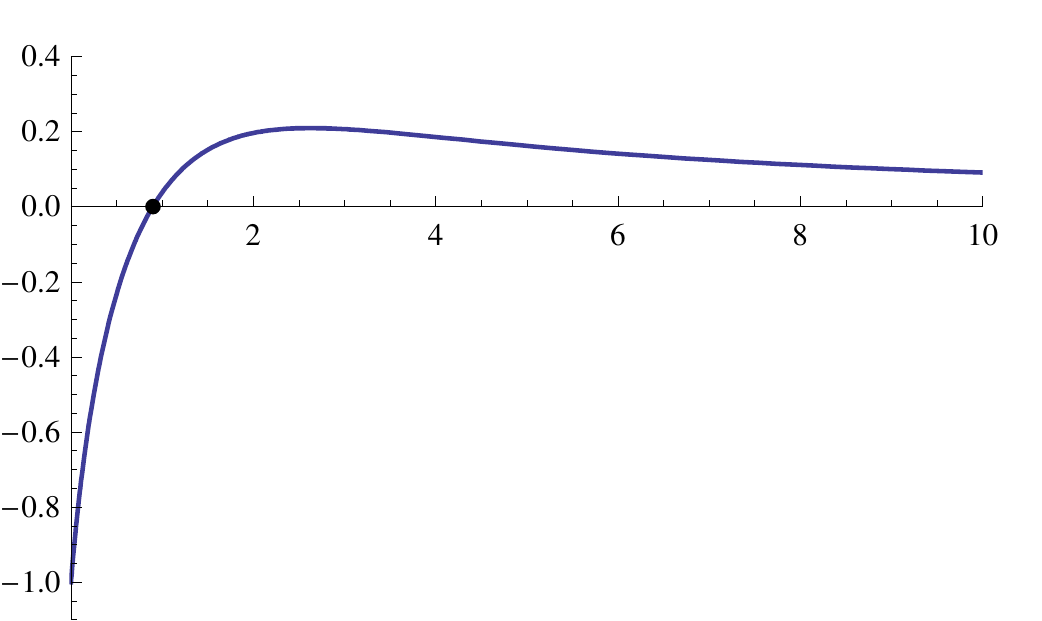}
  \put(-215,120){$f(\gamma)=(1+\gamma)^{-1}-\mmse(X,\gamma)$}
  \put(-5,90){$\gamma$}
  \put(-200,85){$\snr_0$}
  \caption{An example of the difference between the MMSE for standard
    Gaussian input and that of a binary input equally likely to be
    $\pm\sqrt{2}$.  The difference crosses the horizontal axis only
    once.}
  \label{f:diff}
\end{figure}

\begin{IEEEproof} 
  The last of the three statements, $\liminfty{\gamma} f(\gamma) = 0$ 
  always holds because of \prettyref{pr:ub}.
  
  If $\var{X}\le1$, then $f(\gamma)\ge0$ at all $\gamma$ due to \prettyref{pr:ge},
  so that the proposition holds.  We suppose in the following
  $\var{X}>1$.
  An instance of the function $f(\gamma)$ with $X$ equally likely to
  be $\pm \sqrt{2}$ is shown in Fig.~\ref{f:diff}.  Evidently $f(0) =
  1 - \var{X} < 0$.
  Consider the derivative of the difference~\eqref{eq:fg} at any
  $\gamma$ with $f(\gamma)<0$, which by \prettyref{pr:ds}, can be
  written as
  \begin{align}
    f'(\gamma)
    &= \expect{ M_2^2 } - (1 + \gamma)^{-2}
    \label{eq:fgd} \\
    &> \expect{ M_2^2 } - \left( \mmse(X,\gamma) \right)^2
    \label{eq:fge} \\
    &= \expect{ M_2^2 } - \left( \Exp{M_2}
    \right)^2
    \label{eq:fgm} \\
    &\geq 0 \label{eq:fgm0}
  \end{align}
  where~\eqref{eq:fgm} is due to~\eqref{eq:em2}, and~\eqref{eq:fgm0} is due
  to Jensen's inequality.  That is, $f'(\gamma)>0$ as long as
  $f(\gamma)<0$, i.e., the function $f$ can only be strictly
  increasing at every point it is strictly negative.  This further
  implies that if $f(\snr_0)=0$ for some $\snr_0$, the function $f$,
  which is smooth, cannot dip to below zero for any $\gamma>\snr_0$.
  Therefore, the function $f$ has no more than one zero crossing.

  

  
  For any $\sigma$, the above arguments can be repeated with
  $\sigma^2\gamma$ treated as the SNR.  
  It is straightforward to show that the proposition holds with the
  standard Gaussian MMSE replaced by the MMSE of a Gaussian variable
  with variance $\sigma^2$.
\end{IEEEproof}

The single-crossing property can be generalized to the conditional
MMSE defined in~\eqref{eq:mmseu}.\footnote{The single-crossing
  property has also been extended to the parallel degraded MIMO
  scenario~\cite{BusPay}.}

\begin{proposition} \label{pr:fu} %
  Let $X$ and $U$ be jointly distributed variables.  All statements in
  \prettyref{pr:uniq} hold literally if the function $f(\cdot)$ is
  replaced by
  \begin{equation}    \label{eq:fu}
    f(\gamma) = (1+\gamma)^{-1} - \mmse(X,\gamma|U)\ .
  \end{equation}
\end{proposition}

\begin{IEEEproof} 
  For every $u$, let $X_u$ denote a random variable indexed by $u$
  with distribution $P_{X|U=u}$.  Define also a random variable for
  every $u$,
  \begin{align}
    M(u,\gamma) &= M_2(X_u,\gamma) \\
    &= \var{X_u|\sqrt{\snr}\,X_u+N}
  \end{align}
  where $N\sim \mathcal{N} (0,1)$.
  Evidently, $\expect{M(u,\gamma)} = \mmse(X_u,\gamma)$ and hence
  \begin{align}
    f(\gamma)
    &= \frac1{1+\gamma} - \expect{\expcnd{M(U,\gamma)}{U} } \\
    &= \frac1{1+\gamma} - \expect{ M(U,\gamma) }. \label{eq:fgz}
  \end{align}
  Clearly,
  \begin{align}
    f'(\gamma)
    &= - \frac1{(1+\gamma)^2} - \expect{ \pd{\gamma} M(U,\gamma) } \\
    &= \expect{ M^2(U,\gamma) } - \frac1{(1+\gamma)^2} \label{eq:fpg}
  \end{align}
  by \prettyref{pr:ds}.
  In view of~\eqref{eq:fgz}, for all $\gamma$ such that $f(\gamma)<0$,
  we have
  \begin{align}
    f'(\gamma)
     &> \expect{ M^2(U,\gamma) } -
    \left( \expect{ M(U,\gamma) } \right)^2 \\
    &\geq 0
  \end{align}
  by~\eqref{eq:fpg} and Jensen's inequality.  The remaining argument
  is essentially the same as in the proof of \prettyref{pr:uniq}.
\end{IEEEproof}

\subsection{The High-SNR Asymptotics}

The asymptotics of $\mmse(X,\gamma)$ as $\gamma \rightarrow \infty$
can be further characterized as follows.  It is upper
bounded by $1/\gamma$
due to Propositions~\ref{pr:ub} and~\ref{pr:ge}.
Moreover, the MMSE can vanish faster than exponentially in
$\gamma$ with arbitrary rate, under for instance a sufficiently skewed
binary input~\cite{Guo04PhD}.\footnote{In case the input is equally
  likely to be $\pm1$, the MMSE decays as $e^{-\frac12 \snr}$, not
  $e^{-2\snr}$ as stated in~\cite{Guo04PhD, GuoSha05IT}.} On the
other hand, the decay of the MMSE of a non-Gaussian random variable
need not be faster than the MMSE of a Gaussian variable.  For example,
let $X = Z + \sqrt{\sigma_X^2-1}\, B$ where $\sigma_X>1$, $Z\sim
\mathcal{N} (0,1)$ and the Bernoulli variable $B$ are independent.
Clearly, $X$ is harder to estimate than $Z$ but no harder than
$\sigma_X Z$, i.e.,
\begin{equation}  \label{eq:xzb}
  \frac1{1+\gamma} < \mmse(X,\gamma) < \frac{\sigma_X^2}{1+\sigma_X^2\gamma}
\end{equation}
where the difference between the upper and lower bounds is
$\mathcal{O} \left( \gamma^{-2} \right)$.  As a consequence, the
function $f$ defined in~\eqref{eq:fg} may not have any zero even if
$f(0) = 1 - \sigma_X^2 < 0$ and $\liminfty{\gamma} f(\gamma)=0$.
A meticulous study of the high-SNR asymptotics of the MMSE is found
in~\cite{WuVer10ISIT}, where the limit of the product
$\snr\cdot\mmse(X,\snr)$, called the {\em MMSE dimension}, has been
determined for input distributions without singular components.

\section{Applications to Channel Capacity}
\label{s:app}

\subsection{Secrecy Capacity of the Gaussian Wiretap Channel}
\label{s:sc}

This section makes use of the MMSE as an instrument to show that the
secrecy capacity of the Gaussian wiretap channel is achieved by
Gaussian inputs.
The wiretap channel was introduced by Wyner in~\cite{Wyner75BSTJ} in
the context of discrete memoryless channels.  Let $X$ denote the
input, and let $Y$ and $Z$ denote the output of the main channel and
the wiretapper's channel respectively.  The problem is to find the rate
at which
reliable communication is possible through the main channel, while 
keeping the mutual information between the message and the
wiretapper's observation as small as possible.
Assuming that the wiretapper sees a degraded output of the main channel,
Wyner showed that secure communication can achieve any rate
up to the secrecy capacity
\begin{equation}  \label{eq:Cs}
  C_s = \max_X [ I(X;Y) - I(X;Z) ]
\end{equation}
where the supremum is taken over all admissible choices of the input
distribution.  Wyner also derived the achievable rate-equivocation
region.

We consider the following Gaussian wiretap channel studied
in~\cite{LeuHel78IT}:
\begin{subequations} \label{eq:wiretap}
\begin{align}
  Y &= \sqrt{\snr_1}\, X + N_1   \label{eq:wty}  \\
  Z &= \sqrt{\snr_2}\, X + N_2   \label{eq:wtz}
\end{align}  
\end{subequations}
where $\snr_1\geq \snr_2$ and $N_1, N_2 \sim \mathcal{N}(0,1)$ are
independent.  Let the energy of every codeword of length $n$ be
constrained by $\frac1{n} \sum^n_{i=1} x_i^2 \leq 1$.
Reference~\cite{LeuHel78IT}
showed that the optimal input which achieves the supremum in
\eqref{eq:Cs} is standard Gaussian
and that the secrecy capacity is
\begin{equation}
  C_s = \frac12 \log\left( \frac{ 1 + \snr_1 }{ 1 + \snr_2 } \right).
  \label{eq:CsG}
\end{equation}

In contrast to~\cite{LeuHel78IT} which appeals to Shannon's EPI, we
proceed to give a simple proof of the same result
using~\eqref{eq:iei}, which enables us to write for any $X$:
\begin{align}  
  I(X;Y) - I(X;Z)
  &= \frac12 \int_{\snr_2}^{\snr_1}
  \mmse(X,\gamma) \diff \gamma \ . \label{eq:csii2}
\end{align}
Under the constraint $\expect{X^2} \le 1$, the maximum
of~\eqref{eq:csii2} over $X$ is achieved by standard Gaussian input
because it maximizes the MMSE for every SNR under the power
constraint.  Plugging $\mmse(X,\gamma) = (1+\gamma)^{-1}$
into~\eqref{eq:csii2} yields the secrecy capacity given in
\eqref{eq:CsG}.  In fact the whole rate-equivocation region can be
obtained using the same techniques.  Note that the MIMO wiretap
channel can be treated similarly~\cite{BusLiu09EURASIP}.

\subsection{The Gaussian Broadcast Channel}
\label{s:bc}

In this section, we use the single-crossing property to show that
Gaussian input achieves the capacity region of scalar Gaussian
broadcast channels.
Consider a degraded Gaussian broadcast channel also described by the
same model~\eqref{eq:wiretap}.
Note that the formulation of the Gaussian broadcast channel is
statistically identical to that of the Gaussian wiretap channel,
except for a different goal: The rates between the sender and both
receivers are to be maximized, rather than minimizing the rate between
the sender and the (degraded) wiretapper.  The capacity region of
degraded broadcast channels under a unit input power constraint is
given by~\cite{Gallag74PPI}:
\begin{equation}  \label{eq:Cdbc}
  \bigcup_{P_{UX}:\, \expect{X^2} \leq 1} 
  \begin{Bmatrix}
    R_1 \leq I(X;Y|U) \\
    R_2 \leq I(U;Z)
  \end{Bmatrix}
\end{equation}
where $U$ is an auxiliary random variable with $U$--$X$--$(Y,Z)$ being
a Markov chain.
It has long been recognized that Gaussian $P_{UX}$ with standard
Gaussian marginals and correlation coefficient $\expect{UX} =
\sqrt{1-\alpha}$ achieves the capacity.  The resulting capacity region
of the Gaussian broadcast channel is
\begin{equation}  \label{eq:C}
  \displaystyle    \bigcup_{\alpha\in[0,1]} 
  \begin{Bmatrix}
    \displaystyle      R_1 \leq \frac12 \log \big( 1 + \alpha \, \snr_1 \big) \\
    \displaystyle      R_2 \leq \frac12 \logpr{ 
      \frac{ 1 + \snr_2 }{ 1+\alpha\,\snr_2 } }
  \end{Bmatrix}\ .
\end{equation}

The conventional proof of the optimality of Gaussian inputs relies on
the EPI in conjunction with Fano's inequality~\cite{Bergma74IT}.  The
converse can also be proved directly from~\eqref{eq:Cdbc} using only
the EPI~\cite{ElGKim10X, TunSha07ITA}.  In the following we show a
simple alternative proof using the single-crossing property of MMSE.

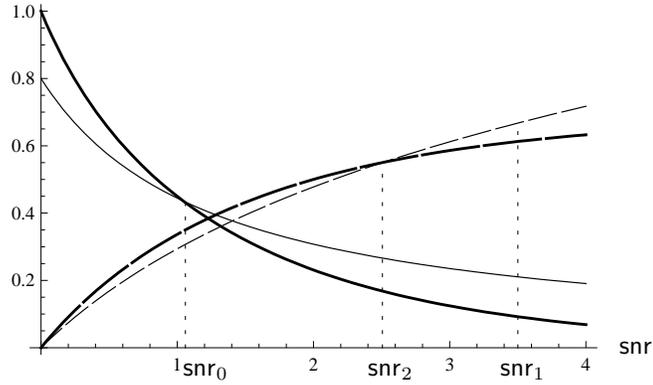
\begin{figure}[htbp]
  \small
  \centering
  \includegraphics[width=\myfigwidth]{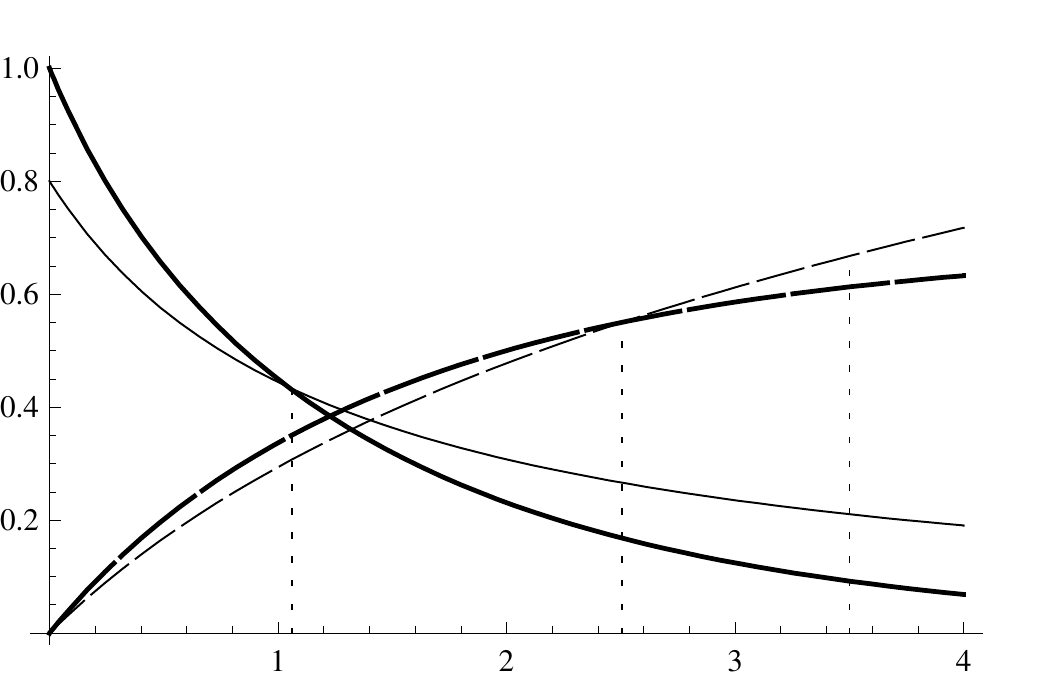}
  \put(-50,0){$\snr_1$}
  \put(-100,0){$\snr_2$}
  \put(-170,0){$\snr_0$}
  \put(-5,8){$\snr$}
  \caption{
    The thin curves show the MMSE (solid line) and mutual information
    (dashed line) of a Gaussian input.  The thick curves show the MMSE
    (solid) and mutual information (dashed) of binary input.  The two
    mutual informations are identical at $\snr_2$, which must be
    greater than $\snr_0$ where the two MMSE curves cross.}
  \label{f:bc}
\end{figure}

Due to the power constraint on $X$, there must exist $\alpha \in
[0,1]$ (dependent on the distribution of $X$) such that
  \begin{align} \label{eq:xy2}
    I(X;Z|U)
    &= \frac12 \logpr{ 1 + \alpha\,\snr_2 } \\
    &= \frac12 \int_0^{\snr_2} \!\! \frac{\alpha}{\alpha \gamma + 1} \diff \gamma.
    \label{eq:xy22}
  \end{align}
  By the chain rule,
  \begin{align}
    I(U;Z) & = I(U,X;Z) - I(X;Z|U) \\
    &= I(X;Z) - I(X;Z|U).
  \end{align}
  By~\eqref{eq:Cdbc} and~\eqref{eq:xy2}, the desired bound on $R_2$ is
  established:
  \begin{align}
    R_2
    &\leq \frac12 \logpr{ 1 + \snr_2 }
    - \frac12 \logpr{ 1 + \alpha\,\snr_2 } \\
    &= \frac12 \logpr{ \frac{ 1 + \snr_2 }{ 1+\alpha\,\snr_2 } }.
  \end{align}

  It remains to establish the desired bound for $R_1$.  The idea is
  illustrated in Fig.~\ref{f:bc}, where crossing of the MMSE curves
  imply some ordering of the corresponding mutual informations.  Note
  that
  \begin{equation}
    I(X;Z|U=u) = \frac12 \int_0^{\snr_2} \mmse(X_u,\gamma) \diff \gamma
  \end{equation}
  and hence
  \begin{equation} \label{eq:xye} %
    I(X;Z|U) = \frac12 \int_0^{\snr_2} \expect{ \mmse(X_U,\gamma|U) }
    \diff \gamma.
  \end{equation}
  Comparing~\eqref{eq:xye} with~\eqref{eq:xy22}, there must exist
  $0\leq \snr_0 \leq \snr_2$ such that
  \begin{equation}
    \expect{ \mmse(X_U,\snr_0|U) } = \frac{\alpha}{\alpha \snr_0 + 1}.
  \end{equation}
  By \prettyref{pr:fu}, 
  this implies
  that for all $\gamma \geq \snr_2 \geq \snr_0$,
  \begin{equation}  \label{eq:exug}
    \expect{ \mmse(X_U,\gamma|U) } \leq \frac{\alpha}{\alpha \gamma + 1}.
  \end{equation}
  Consequently,
  \begin{align}
    R_1
    &\leq I(X;Y|U) \\
    &= \frac12 \int_0^{\snr_1} \expect{ \mmse(X_U,\gamma|U) } \diff \gamma \\
    &= \frac12 \bigg( \int_0^{\snr_2} + \int_{\snr_2}^{\snr_1} \bigg)
    \expect{ \mmse(X_U,\gamma|U) } \diff \gamma \\
    &\leq \frac12 \logpr{ 1 + \alpha\,\snr_2 }
    + \frac12 \int_{\snr_2}^{\snr_1}
    \frac{\alpha}{\alpha \gamma + 1} \diff \gamma \label{eq:ineq} \\
    &= \frac12 \logpr{ 1 + \alpha\,\snr_1 }
  \end{align}
  where the inequality~\eqref{eq:ineq} is due to~\eqref{eq:xy2},
  \eqref{eq:xye} and~\eqref{eq:exug}.

\subsection{Proof of a Special Case of EPI}
\label{s:epi}

As another simple application of the single-crossing property, we show
in the following that
\begin{equation}  \label{eq:epig}
  e^{2h(X+Z)} \ge e^{2h(X)} + 2\pi e \sigma^2_Z 
\end{equation}
for any independent $X$ and $Z$ as long as the differential entropy of
$X$ is well-defined and $Z$ is Gaussian with variance $\sigma_Z^2$.
This is in fact a special case of Shannon's entropy power inequality.
Let $W\sim\mathcal{N}(0,1)$ and $a^2$ be the ratio of the entropy
powers of $X$ and $W$, so that
\begin{equation} \label{eq:hxaz} %
  h(X)=h(aW)=\frac12\log\left( 2\pi e a^2 \right) .
\end{equation}
Consider the difference
\begin{equation}  \label{eq:ha}
  \begin{split}
    h&\left( \sqrt{\snr}\,X + N \right) - h\left( \sqrt{\snr}\,aW + N \right)\\
    &= \frac12 \int_0^\snr \mmse(X,\gamma) - \mmse(aW,\gamma) \diff
    \gamma
  \end{split}
\end{equation}
where $N$ is standard Gaussian independent of $X$ and $W$.  In the
limit of $\snr\to\infty$, the left hand side of~\eqref{eq:ha} vanishes
due to~\eqref{eq:hxaz}.  By \prettyref{pr:uniq}, the integrand
in~\eqref{eq:ha} as a function of $\gamma$ crosses zero only once,
which implies that the integrand is initially positive, and then
becomes negative after the zero crossing (cf.~Fig.~\ref{f:diff}).
Consequently, the integral~\eqref{eq:ha} is positive and increasing
for small $\snr$, and starts to monotonically decrease after the zero
crossing.  If the integral crosses zero it will not be able to cross
zero again.  Hence the integral in~\eqref{eq:ha} must remain positive
for all $\snr$ (otherwise it has to be strictly negative as
$\snr\to\infty$).  Therefore,
\begin{align}
  \expb{ 2 h\left( \sqrt{\snr} X + N \right) }
  &\ge \expb{ h\left( \sqrt{\snr} W + N \right) } \\
  &= 2 \pi e \left( a^2 \snr + 1 \right) \\
  &= \expb{2h\left(\sqrt{\snr} X\right)} + 2 \pi e
\end{align}
which is equivalent to~\eqref{eq:epig} by choosing
$\snr=\sigma_Z^{-2}$ and appropriate scaling.

The preceding proof technique also applies to conditional EPI, which
concerns $h(X|U)$ and $h(X+Z|U)$, where $Z$ is Gaussian independent of
$U$.  The conditional EPI can be used to establish the capacity region
of the scalar broadcast channel in~\cite{Bergma74IT, ElGKim10X}.


\section{Concluding Remarks}
\label{s:con}

This paper has established a number of basic properties of the MMSE in
Gaussian noise as a transform of the input distribution and function
of the SNR.  Because of the intimate relationship MMSE has with
information measures, its properties find direct use in a number of
problems in information theory.

The MMSE can be viewed as a transform from the input distribution to a
function of the SNR: $P_X \mapsto \{\mmse(P_X,\gamma),\;
\gamma\in[0,\infty)\}$.  An interesting question remains to be
answered: Is this transform one-to-one?  We have the following
conjecture:

\begin{conjecture}
  For any zero-mean random variables $X$ and $Z$, $\mmse(X,\snr)
  \equiv \mmse(Z,\snr)$ for all $\snr \in [0,\infty)$ if and only if
  $X$ is identically distributed as either $Z$ or $-Z$.
  \label{cj:ee}
\end{conjecture}


There is an intimate relationship between the real analyticity of MMSE
and \prettyref{cj:ee}.  In particular, MMSE being real-analytic at
zero SNR for all input and MMSE being an injective transform on the
set of all random variables (with shift and reflection identified)
cannot both hold.  This is because given the real analyticity at zero
SNR,
MMSE can be extended to an open disk $D$ centered at zero via the power
series expansion, where the coefficients depend only on the moments of
$X$.  Since solution to the Hamburger moment problem is not unique in
general, there may exist different $X$ and $X'$ with the same moments,
and hence their MMSE function coincide in $D$.  By the identity
theorem of analytic functions, they coincide everywhere, hence on the
real line.  Nonetheless, if one is restricted to the class of
sub-Gaussian random variables, the moments determine the distribution
uniquely by Carleman's condition~\cite{Feller71}.




\appendices

\section{Proof of \prettyref{pr:mmt}}
\label{a:mmt}

\begin{IEEEproof}
  Let $Y=\sqrt{\snr}\,X+N$ with $\snr>0$.  Using~\eqref{eq:XN} and
  then Jensen's inequality twice, we have
  \begin{align}
    &\expect{ | X - \expcnd{X}{Y} |^n } \nonumber \\
    &\qquad= \snr^{-\frac{n}2} 2^n\,\expect{ 2^{-n}| \expcnd{N}{Y} - N |^n } \\
    &\qquad\le \snr^{-\frac{n}2} 2^{n-1}
    \, \expect{ | \expcnd{N}{Y} |^n + | N |^n } \\
    &\qquad\le \snr^{-\frac{n}2} \, 2^n\, \expect{ |N|^n }
    \label{eq:ENn}
  \end{align}
  which leads to~\eqref{eq:mmt} because
  \begin{align}
    \expect{|N|^n} &= \sqrt{\frac{2^n}{\pi}}\Gamma\left(\frac{n+1}{2}\right) \\
    &\le \sqrt{n!} \ .  \label{eq:n!}
  \end{align}
\end{IEEEproof}

\section{Proof of \prettyref{pr:subg.post}}
\label{a:subg.post}

\begin{IEEEproof}
  We use the characterization by moment generating function
  in~\prettyref{lmm:subg.char}:
  \begin{align}
    \expect{ e^{t X_y} } %
    &= \frac1{h_0(y;a)} \expect{ e^{tX} \varphi(y-aX) } \\
    &= \frac{\varphi(y)}{h_0(y; a)} \expect{\expb{(t+ a y) X
        - \frac{a^2 X^2}{2}}}    \\
    &\leq \frac{\varphi(y)}{h_0(y; a)} \expb{\frac{(t+ a y)^2}{2a^2}}    \label{eq:Xu.MGF}\\
    &\leq \frac{\varphi(y)}{h_0(y; a)} \expb{\frac{t^2}{a^2} + y^2}
    \label{eq:Xu.MGF.1}
\end{align}
where~\eqref{eq:Xu.MGF} and~\eqref{eq:Xu.MGF.1} are due to elementary
inequalities.  Using Chernoff's bound and~\eqref{eq:Xu.MGF.1}, we have
\begin{align}
  \Probk{X_y \geq x}
  &\leq \expect{e^{t (X_y-x)}}    \\
  &\leq \frac{\varphi(y) e^{y^2}}{h_0(y; a)} \expb{\frac{t^2}{a^2} -t x}
\end{align}
for all $x, t > 0$.
Choosing $t = \frac{a^2 x}{2}$ yields
\begin{equation}
  \Probk{X_y \geq x} \leq  \frac{e^{\frac{y^2}{2}}}{ h_0(y; a)} 
  \varphi\pth{\frac{ax}{\sqrt{2}}}.
\end{equation}
Similarly, $\Probk{X_y \leq - x}$ admits the same bound as above,
and~\eqref{eq:Xy.Q} follows from the union bound. Then, using an
alternative formula for moments~\cite[p.~319]{Stirza03}:
\begin{align}
  \expect{|X_y|^n}
  &= n \int_0^\infty x^{n-1} \Probk{|X_y| \geq x} \diff x    \\
  &\leq \frac{2n e^{\frac{y^2}{2}}}{h_0(y; a)} \int_0^\infty
  x^{n-1} \varphi\pth{\frac{ax}{\sqrt{2}}} \diff x    \label{eq:fa.1}\\
  &\leq \frac{n e^{\frac{y^2}{2}}}{h_0(y; a)}
  \left(\frac{\sqrt{2}}{|a|} \right)^n \expect{ |N|^{n-1} }
  \label{eq:fa.4}
\end{align}   
where $N\sim\mathcal{N} (0,1)$ and~\eqref{eq:fa.1} is due
to~\eqref{eq:Xy.Q}.  The inequality~\eqref{eq:Xyn} is thus established
by also noting~\eqref{eq:n!}.

  Conditioned on $Y=y$, using similar techniques leading
  to~\eqref{eq:ENn}, we have
  \begin{align}
    & \expcnd{ | X - \expcnd{X}{Y} |^n }{ Y=y } \nonumber\\
    &\quad \le 2^{n-1} (\expcnd{|X|^n}{ Y=y } +  |\expcnd{X}{Y = y}|^n ) \label{eq:XnY} \\
    &\quad\le 2^{n} \expcnd{|X|^n}{Y=y}     \label{eq:XnY.1} 
  \end{align}
  which is~\eqref{eq:mmty}.
\end{IEEEproof}

\section{Proof of \prettyref{lmm:gi}}
\label{a:gi}

We first make the following observation:

\begin{lemma}
  For every $i=0,1,\dots$, the function $g_i$ is a finite weighted sum
  of functions of the following form:
  \begin{equation}
    \frac1{h_0^{k-1}} \prod_{j=1}^k h_{n_j}^{(m_j)}
    \label{eq:gi.1}
  \end{equation}
  where $n_j, m_j, k=0,1,\dots$.
  \label{lmm:gi.form}
\end{lemma}

\begin{IEEEproof}
  We proceed by induction on $i$: The lemma holds for $i=0$ by
  definition of $g_0$.
Assume the induction hypothesis holds for $i$. Then
\begin{equation}
  \begin{split}
    \fracp{}{a} \left(\frac1{h_0^{k-1}} \prod_{j=1}^k
      h_{n_j}^{(m_j)}\right) &= \frac{-(k-1)}{h_0^k} h_0'
    \prod_{j=1}^k h_{n_j}^{(m_j)} \\
    + \frac1{h_0^{k-1}} & \sum_{l=1}^k h_{n_l}^{(m_l+1)} \prod_{j \neq
      l} h_{n_j}^{(m_j)}
  \end{split}
\end{equation}
which proves the lemma.
\end{IEEEproof}  

To show the absolutely integrability of $g_i$, it suffices to show the
function in~\eqref{eq:gi.1} is integrable:
\begin{align}
  \int^\infty_{-\infty} &\Bigg| \frac{1}{h_0^{k-1}(y;a)} \prod_{j=1}^k
  \fracpk{h_{n_j}(y;a)}{a}{m_j} \Bigg| \diff y \nonumber \\
  = & ~ \expect{\prod_{j=1}^k \left| \frac1{h_0(Y;a)}
      \fracpk{h_{n_j}}{a}{m_j}(Y; a) \right| }	\label{eq:int.1} \\
  = & ~ \expect{\prod_{j=1}^k \left| \expcnd{
      X^{n_j + m_j} H_{m_j}(Y-aX) }{ Y } \right| } \label{eq:int.2}\\
  \leq & ~ \prod_{j=1}^k \left[\expect{ \left( \expcnd{ | X^{n_j + m_j}
          H_{m_j}(Y-aX) | }{Y} \right)^k}
  \right]^{\frac{1}{k}} \label{eq:int.3}\\
  \leq & ~ \prod_{j=1}^k \left[\expect{|X|^{k (n_j + m_j)}} \expect{|H_{m_j}(N)|^k}\right]^{\frac{1}{k}} \label{eq:int.4}\\
  < &~ \infty \label{eq:int.5}
\end{align}
where~\eqref{eq:int.2} is by~\eqref{eq:hi.exp},~\eqref{eq:int.3} is by
the generalized H\"older inequality~\cite[p.~46]{LieLos01},
and~\eqref{eq:int.4} is due to Jensen's inequality and the
independence of $X$ and $N=Y-aX$.

\section{Proof of \prettyref{pr:mmse.ana} on the Analyticity}
\label{a:ana}


We first assume that $X$ is sub-Gaussian.

Note that $\varphi$ is real analytic everywhere with infinite radius
of convergence, because $\varphi^{(n)}(y) = (-1)^n H_n(y) \varphi(y)$
and Hermite polynomials admits the following
bound~\cite[p.~997]{table.integrals}:
\begin{equation}
	|H_n(y)| \leq \kappa \sqrt{n!} \,e^{\frac{y^2}{4}}
	\label{eq:Hn.ub}
\end{equation}
where $\kappa$ is an absolute constant. Hence 
\begin{equation}
\lim_{n \to \infty} \left|\frac{\varphi^{(n)}(y)}{n!}\right|^{\frac{1}{n}} = 0	
\end{equation}
and the radius of convergence is infinite at all $y$. Then
\begin{equation}
\varphi(y - a' x) = \sum_{n = 0}^{\infty} \frac{H_n(y - a x) \varphi(y- a x) x^n}{n!} (a'-a)^n 
	\label{eq:phi.series}
\end{equation}
holds for all $a, x \in \reals$.  
By \prettyref{lmm:subg.char}, there exists $c > 0$, such that 
$\expect{|X|^{n}} \leq c^n \sqrt{n!}$ for all $n=1,2,\dots$.
By~\eqref{eq:Hn.ub}, it is easy to see that $|H_n(y)\varphi(y)|\le
\kappa \sqrt{n!}$ for every $y$.  Hence
\begin{equation}
  \expect{\left|H_n(y -a X) \varphi(y-aX) X^n\right|}
   \leq \kappa c^n n!\,. 
\end{equation}
Thus for every $|a' - a| < R \triangleq \frac{1}{c}$,
\begin{equation}
  \sum_{n =0}^{\infty}  \frac{|a'-a|^n}{n!} 
  \expect{\left| (H_n \cdot \varphi)(y - a X) X^n\right|}< \infty.
\end{equation}
Applying Fubini's theorem to~\eqref{eq:phi.series} yields
\begin{equation}
h_0(y;a') = \sum_{n = 0}^{\infty}  \frac{(a'-a)^n}{n!} \expect{(H_n \cdot \varphi)(y - a X) X^n}
	\label{eq:h0.series}
\end{equation}
Therefore, $h_0(y;a)$ is real analytic at $a$ and the radius of convergence is lower bounded by $R$ independent of $y$. Similar conclusions also apply to $h_1(y; a)$ and
\begin{equation}
	h_1(y;a') = \sum_{n = 0}^{\infty} \frac{(a'-a)^n}{n!} \expect{(H_n \cdot \varphi)(y - a X) X^{n+1}}
	\label{eq:h1.series}
\end{equation}
holds for all $y \in \reals$ and all $|a'-a| < R$. 
Extend $h_0(y;a)$ and $h_1(y;a)$ to the complex disk $D(a,R)$ by the power series~\eqref{eq:h0.series} and~\eqref{eq:h1.series}. 
By~\eqref{eq:h0.nonz}, there exists $0 < r < R/2$, such that $h_0(y; z)$ does not vanishes on the disk $D(a,r)$. By~\cite[Proposition 1.1.5]{KraPar02}, for all $y \in \reals$,
\begin{equation}
  g_0(y; z) = \frac{h_1^2(y;z)}{h_0(y;z)}
\end{equation}
is analytic in $z$ on $D(a,r)$. 

By assumption~\eqref{eq:h0lb}, there exist $B, c > 0$, such that
\begin{equation}
	|h_0(y; z)| \geq c \, h_0(y; \Rerm{z})
	\label{eq:h0_tau}
\end{equation}
for all $z \in D(a, r)$ and all $|y| \geq B$. Define
\begin{equation}
	m_0^{B}(z) = \int_{-B}^B g_0(y; z) \diff y.
\end{equation}
Since $(y,z) \mapsto g_0(y;z)$ is continuous, 
for every closed curve $\gamma$ in $D(a,r)$, we have $\oint_{\gamma} \int_{-B}^B |g_0(y; z)| \diff y \diff z < \infty$. By Fubini's theorem, 
\begin{align}
  \oint_{\gamma} \int_{-B}^B g_0(y; a) \diff y \diff z = \int_{-B}^B
  \oint_{\gamma} g_0(y; a) \diff z \diff y = 0
\end{align}
where the last equality follows from the analyticity of
$g_0(y;\cdot)$. By Morera's theorem~\cite[Theorem 3.1.4]{GreKra06},
$m_0^{B}$ is analytic on $D(a,r)$.

Next we show that as $B \to \infty$, $m_0^{B}$ tends to $m_0$ uniformly in $z \in D(a,r)$. Since uniform limit of analytic functions is analytic~\cite[p. 156]{lang.complex}, we obtain the analyticity of $m_0$. To this end, it is sufficient to show that $\{|g_0(\cdot\,;z)|: z \in D(a,r)\}$ is uniformly integrable. Let $z = s + i t$. Then
\begin{align}
|h_1(y;z)|
= & ~ |\expect{X \varphi(y - z X)}|	\\
\leq & ~ \expect{|X| |\varphi(y -z X)|}	\\
= & ~ \expect{|X| \varphi(y - s X) e^{\frac12 t^2 X^2}} \label{eq:h1.Re}.
\end{align}
Therefore, for all $z \in D(a,r)$,
\begin{align}
  \int_{\reals} & |g_0(y; z)|^2 \diff y - \int_{-K}^K |g_0(y; z)|^2
  \diff y\nonumber \\
  &\leq \frac1{c^2} \int_{\reals} \left|
    \frac{h_1(y; z)}{h_0(y; s)}\right|^4 h_0^2(y; s) \diff y \label{eq:ui.1}\\
  &\leq \frac1{c^2} \int_{\reals} \left|
    \frac{\expect{|X|e^{\frac12t^2X^2}\varphi(y-sX)}}{h_0(y;
      s)}\right|^4 h_0(y; s) \diff y \label{eq:ui.15}\\
  &\leq \frac1{c^2}\expect{ \pth{\expcnd{|X|
        e^{\frac{t^2 X^2}{2}}}{Y_{s^2}}}^4} \label{eq:ui.2}\\
  &\leq \frac1{c^2} \expect{X^4 e^{2 r^2 X^2}} \label{eq:ui.3} 
\end{align}
where~\eqref{eq:ui.1} is by~\eqref{eq:h0lb},~\eqref{eq:ui.15} is by
$|h_0(y;s)| \leq 1$,~\eqref{eq:ui.2} is by~\eqref{eq:h1.Re},
and~\eqref{eq:ui.3} is due to Jensen's inequality and $|t| \leq r$.
Since $X$ is sub-Gaussian satisfying~\eqref{eq:subg.cdf} and $r < R/2
= 1/(2c)$,
\begin{align}
  \expect{X^4 e^{2 r^2 X^2}}
  \leq & ~ \sum_{n =0}^{\infty} \frac{(2r^2)^n}{n!}\expect{|X|^{2n+4}} 
  \label{eq:X4.1} \\
  \leq & ~ \sum_{n =0}^{\infty} \frac{(2r^2)^n}{n!} \sqrt{(2n+4)!}~c^{2n+4} \\
  \leq & ~ 4 c^4\sum_{n =0}^{\infty}  (n^2+3n+2) (2r c )^{2n} \\
  < & ~ \infty \ . \label{eq:X4.4}
\end{align}
Therefore $\{|g_0(\cdot \,;z)|: z \in D(a,r)\}$ is $L^2$-bounded,
hence uniformly integrable.  We have thus shown that $m_0(a)$, i.e.,
the MMSE, is real analytic in $a$ on $\reals$.

We next consider positive SNR and drop the assumption of
sub-Gaussianity of $X$.  Let $a_0 > 0$ and fix $\delta$ with $0 <
\sqrt{\delta} < a_0/2$.  We use the incremental-SNR representation for
MMSE in \prettyref{eq:incremental}. Define $\bar{X}_u$ to be
distributed according to $X - \expect{X|Y_{\delta} = u}$ conditioned
on $Y_{\delta} = u$ and recall the definition of and $h_i(y; a|u;
\delta)$ in~\eqref{eq:hibar}. In view of \prettyref{pr:subg.post},
$\bar{X}_u$ is sub-Gaussian whose growth of moments only depends on
$\delta$ (the bounds depend on $u$ but the terms varying with $n$ do
not depend on $u$).  Repeating the arguments from~\eqref{eq:Hn.ub}
to~\eqref{eq:h1.series} with $c = \sqrt{2/\delta}$, we conclude that
$h_0(y; a|u; \delta)$ and $h_1(y; a|u; \delta)$ are analytic in $a$
and the radius of convergence is lower bounded by $R =
\sqrt{\delta/2}$, independent of $u$ and $y$.

Let $r < \sqrt{\delta} / 4$. The remaining argument follows as in the
first part of this proof, except that~\eqref{eq:ui.1}--\eqref{eq:X4.4}
are replaced by the following estimates: Let $\tau = t^2/2$, then
\begin{align}
  & \expect{ \pth{\expcnd{|\bar{X}| e^{\tau X^2}}{Y_{s^2},
        Y_{\delta}}}^4} \nonumber\\
  &\qquad\leq \expect{ \pth{\expcnd{|\bar{X}| 
        e^{\tau X^2}}{Y_{\delta}}}^4} \label{eq:iq.2}\\
  &\qquad= \expect{ \prod_{i=1}^4 \sum_{n_i=0}^{\infty} \frac{\tau^{n_i}}{n_i!} \expcnd{|\bar{X}|^{2n_i+1}}{Y_\delta}} \label{eq:iq.4}\\
  &\qquad\leq \sum_{n_1,n_2,n_3,n_4=0}^{\infty} \pth{\frac{8
      \tau}{\delta}}^{\sum_i n_i+1}
  \binom{\sum_i n_i}{n_1,n_2,n_3,n_4} \label{eq:iq.6}\\
  &\qquad\leq \left(\frac{8\tau}{\delta}\right)
  \sum_{n_1,n_2,n_3,n_4=0}^{\infty}
  \pth{\frac{32 \tau}{\delta^2}}^{\sum_i n_i}\\
  &\qquad= \left(\frac{8\tau}{\delta}\right) \pth{\sum_{n=0}^{\infty}
    \pth{\frac{32 \tau}{\delta^2}}^n}^4 \label{eq:iq.7}\\
  &\qquad< \infty \label{eq:iq.8}
\end{align}
where~\eqref{eq:iq.2} is by Jensen's inequality,~\eqref{eq:iq.4} is by
Fubini's theorem,~\eqref{eq:iq.8} is because $\tau \leq r^2/2 <
\delta^2/32$, and~\eqref{eq:iq.6} is by \prettyref{lmm:bMi}, to be
established next.

Let $M_i$ be defined as in \prettyref{s:dMi}.  The following lemma
bounds the expectation of products of $|M_i|$:
\begin{lemma}
  For any $\snr > 0$, $k, i_j, n_j \in \naturals$,
  \begin{equation}
    \expect{\prod_{j=1}^k |M_{i_j}|^{n_j}} \leq
    \snr^{-\frac{n}2} 2^n \sqrt{n!}
    \label{eq:bMi}
  \end{equation}
  where $n = \sum_{j=1}^k i_j n_j$.
	\label{lmm:bMi}
\end{lemma}

\begin{IEEEproof}
  In view of \prettyref{pr:mmt}, it suffices to establish:
  \begin{align}
    \expect{\prod_{j=1}^k |M_{i_j}|^{n_j}}
    &= \expect{\prod_{j=1}^k \prod_{l=1}^{n_j} |M_{i_j}|} \\
    &\le \prod_{j=1}^k \prod_{l=1}^{n_j} \left( \expect{|M_{i_j}|^{
          \frac{n}{i_j}}} \right)^{\frac{i_j}{n}} \label{eq:Mi.p.2} \\
    &\le \prod_{j=1}^k \prod_{l=1}^{n_j} \left( \expect{ |X -
        \expect{X \mid Y}|^n } \right)^{\frac{i_j}{n}} \label{eq:Mi.p.3} \\
    &= \expect{ |X - \expect{X \mid Y}|^n} \label{eq:Mi.p.4}
\end{align}
where~\eqref{eq:Mi.p.2} and~\eqref{eq:Mi.p.3} are due to the
generalized H\"older's inequality and Jensen's inequality, respectively.
\end{IEEEproof}

\section{Proof of \prettyref{pr:ds} on the Derivatives}
\label{a:ds}

The first derivative of the mutual information with respect to the SNR
is derived in~\cite{GuoSha05IT} using the incremental channel
technique.  The same technique is adequate for the analysis of the
derivatives of various other information theoretic and estimation
theoretic quantities.  

The MMSE of estimating an input with zero mean, unit variance and
finite higher-order moments admits the Taylor series expansion at the
vicinity of zero SNR given by~\eqref{eq:taylor}.
In general, given a random variable $X$ with arbitrary mean and
variance, we denote its central moments by
\begin{equation}
  m_i = \expect{ ( X - \expect{X} )^i }, \quad i=1,2,\dots.
\end{equation}
Suppose all moments of $X$ are finite, the random variable can be
represented as $X= \expect{X} +\sqrt{m_2}\,Z$ where $Z$ has zero mean
and unit variance.  Clearly, $\Exp{Z^i} = m_2^{-\frac{i}{2}} m_i$.  By
\eqref{eq:taylor} and \prettyref{pr:sc},
\begin{align}
  &\mmse(X,\snr) \nonumber \\
  &\quad= m_2  \, \mmse(Z,\snr\,m_2) \\
  &\quad= m_2 - m_2^2\snr + \left(2m_2^3-m_3^2\right)\frac{\snr^2}{2}
  - \big( m_4^2 - 6 m_4 m_2^2  \nonumber\\
  &\qquad\qquad - 12 m_3^2 m_2 + 15 m_2^4 \big)
  \frac{\snr^3}6 + \mathcal{O}\left(\snr^4\right).
\end{align}
In general, taking into account the input variance, we have:
\begin{align}
  \mmse'(X,0) &= -m_2^2   \label{eq:d1} \\
  \mmse''(X,0) &= 2 m_2^3 - m_3^2   \label{eq:d2} \\
  \mmse'''(X,0)&= - m_4^2 + 6 m_4 m_2^2 + 12 m_3^2 m_2 - 15m_2^4\,.
  \label{eq:d3}
\end{align}

Now that the MMSE at an arbitrary SNR is rewritten as the expectation
of MMSEs at zero SNR, we can make use of known derivatives at zero SNR
to obtain derivatives at any SNR.  Let $X_{y;\snr} \sim
P_{X|Y_\snr=y}$.  Because of~\eqref{eq:d1}, 
\begin{equation} \label{eq:dv2y} %
  \frac{\diff \mmse( X_{y;\snr}, \gamma ) 
  }{\diff \gamma}
  \Big|_{\gamma=0^+} = - \left( \var{X|Y_\snr=y} \right)^2 \ .
\end{equation}
Thus,
\begin{align}
  \frac{\diff \mmse(X,\snr)}{\diff \snr}
  &= \pd{\gamma} \mmse(X,\snr+\gamma) \Big|_{\gamma=0^+} \\
  &= \pd{\gamma} \mmse( X, \gamma | Y_\snr )
  \Big|_{\gamma=0^+} \label{eq:dg0} \\
  &= - \expect{ \left( \var{X|Y_\snr} \right)^2 } \label{eq:v2} \\
  &= - \expect{ M_2^2 }
\end{align}
where~\eqref{eq:dg0} is due to \prettyref{pr:ic} and the fact that the
distribution of $Y_\snr$ is not dependent on $\gamma$, and
\eqref{eq:v2} is due to~\eqref{eq:dv2y} and averaging over $y$
according to the distribution of $Y_\snr=\sqrt{\snr}\,X+N$.  Hence
\eqref{eq:ds1} is proved.  Moreover, because of~\eqref{eq:d2},
\begin{equation} \label{eq:dv3}
  \begin{split}
    & \frac{\diff^2 \mmse( X_{y;\snr}, \gamma ) 
    }{\diff \gamma^2}
    \Big|_{\gamma=0} = 2 \, \left( \var{X|Y_\snr=y} \right)^3  \\
    &\qquad \qquad \quad - \left(
      \expcnd{(X - \expcnd{X}{Y_\snr})^3}{Y_\snr=y} \right)^2
  \end{split}
\end{equation}
which leads to~\eqref{eq:ds2} after averaging over the distribution of
$Y_\snr$.  Similar arguments, together with~\eqref{eq:d3}, lead to the
third derivative of the MMSE which is obtained as~\eqref{eq:ds3}.

\section*{Acknowledgement}

The authors would like to thank the anonymous reviewers for their
comments, which have helped to improve the paper noticeably.  The
authors would also like to thank Miquel Payar\'o, Daniel Palomar and
Ronit Bustin for their comments.



\end{document}